    \documentclass[twocolumn,twocolappendix]{aastex7}
    \usepackage{
        amsmath,
        amssymb,
        newtxtext,
        newtxmath,
        ae, aecompl,
        graphicx,
        booktabs,
        xcolor,
    }

    \newcommand*{\scinot}[2]{#1\times10^{#2}}
    \newcommand*{\rd}[2]{\frac{\mathrm{d}#1}{\mathrm{d}#2}}

    \newcommand*{\abs}[1]{\left|#1\right|}
    
    \newcommand*{\p}[1]{\left(#1\right)}
    \newcommand*{\s}[1]{\left[#1\right]}
    \newcommand*{\z}[1]{\left\{#1\right\}}
    \let\bm\undefined
    \newcommand*{\bm}[1]{\mathbf{#1}}
    \newcommand*{\uv}[1]{\hat{\boldsymbol{\mathbf{#1}}}}

    \newlength{\colummwidth}
    \usepackage[T1]{fontenc}

    \colorlet{Corr}{red}
    \received{XXXX}
    \revised{YYYY}
    \accepted{ZZZZ}
    \submitjournal{ApJ}

\shorttitle{Stellar Obliquities and Disks}
\shortauthors{Y.\ Su and D.\ Lai}

\begin{document}

\title{Stellar Obliquity Excitation via Disk Dispersal-Driven Resonances in
Binaries}

\author[orcid=0000-0001-8283-3425]{Yubo Su}% chktex 8
\affiliation{Department of Astrophysical Sciences, Princeton University, 4 Ivy
Ln, Princeton, NJ 08544, USA}
\email[show]{yubosu@princeton.edu}
\correspondingauthor{Yubo Su}

\author[orcid=0000-0002-1934-6250]{Dong Lai}% chktex 8
\affiliation{Cornell Center for Astrophysics and Planetary Science, Department
of Astronomy, Cornell University, Ithaca, NY 14853, USA}
\affiliation{Tsung-Dao Lee Institute, Shanghai Jiao Tong University, Shanghai
201210, China}
\email{dl57@cornell.edu}

\begin{abstract}

The stellar obliquity of a planetary system is often used to help constrain the
system's formation and evolution.
One of the mechanisms to reorient the stellar spin involves a secular resonance
crossing due to the dissipation of the protoplanetary disk when the system also
has an inclined, distant ($\sim 300\;\mathrm{au}$) binary companion.
This mechanism is likely to operate broadly due to the $\sim 50\%$ binary
fraction of FGK dwarfs and can play an important role in setting the initial
stellar obliquities prior to any dynamical evolution.
In this work, we revisit this mechanism analytically for idealized, homologously
evolving disk models and show that the resulting stellar obliquities are broadly
distributed between $60^\circ$ and $180^\circ$ for most warm and cold planets.
We further show that non-homologus disk dissipation, such as the development of
a photoevaporatively-opened gap at $\sim 2\;\mathrm{au}$, can help maintain
orbital alignment of warm planets, in agreement with observations.
Our results represent the proper primordial obliquities for planetary systems
with distant binary companions.
They also represent the obliquities of stars with no present-day binary
companions if these companions are dynamically unbound during the birth cluster
phase of evolution, a process that occurs on a comparable timescale as the
disk-driven obliquity excitation.

\end{abstract}


\section{Introduction}\label{s:intro}

In traditional theories of planet formation, the orbits of planets and the spins
of their host stars are expected to be closely aligned, because they form out of
the same protoplanetary disks (PPDs).
This is motivated by the observed small $7^\circ$ misalignment between the solar
spin axis and the normal to the ecliptic \citep{beck2005sunobl}.
Measurement of this misalignment angle in exoplanetary systems, known as the
\emph{stellar obliquity}\footnote{Not to be confused with the ``planetary
obliquity'' of a planet, the misalignment angle between a planet's spin axis and
its own orbit normal.} has become moderately widespread in the recent years
using various techniques such as the Rossiter-Mclaughlin effect
\citep{rossiter1924, mclaughlin1924}.
Surprisingly, a substantial (though still minority) fraction of stellar
obliquities are large, including tentative evidence for an excess near
$90^\circ$ (\citealp{albrecht2021perp}, but see \citealp{dong2023obl,
siegel2023obl}).
The origin of these large misalignments is currently unknown, and a broad range
of causes has been proposed (see \citealp{albrecht2022review} for a good
review).
Distinguishing among these scenarios requires careful studies of various trends
in the measured population.

Among the measured stellar obliquities (we will refer to these as just
``obliquities''), a few notable clues help constrain the mechanisms responsible
for misalignments.
The most well-known trend is the strong dependence of obliquity on stellar
temperature among hot Jupiter hosts: stars with surface temperatures below
$6100\;\mathrm{K}$ have low obliquities, while their hotter counterparts have
large obliquities \citep{winn2010hj_obl}.
This is commonly attributed to dissipation of the tide raised on the star by the
hot Jupiter, a process that is much more efficient in the deep convective
envelopes present in stars cooler than $\sim 6100\;\mathrm{K}$
\citep{albrecht2012hj_tide, lai2012, zanazzi2024_hjlock1}.
However, recent observations of lower-mass \emph{Kepler} systems suggest
preferential spin-orbit alignment below $6100\;\mathrm{K}$ as well
\citep{louden2021, louden2024}; the tides raised by such low-mass planets are
unlikely to be capable of realigning the stellar spin in standard tidal
scenarios, as the alignment timescale is inversely proportional to the planet
mass.
This result is consistent with the general lack of correlation between
spin-orbit misalignment and planet properties reported by
\citet{albrecht2021perp}.
While the evidence is certainly not yet conclusive, it motivates a closer look
at mechanisms for producing spin-orbit misalignment that are largely independent
of planetary mass.
Many mechanisms fit this constraint; a few that we will not discuss include:
magnetic torques \citep{lai2011mag, foucart2011mag, spalding2015magnetic,
spalding2016mag}, stochastically-excited internal gravity waves
\citep{rogers2012igw, rogers2013igw}, stellar spin stochasticity at formation
\citep{bate2010, spalding2014protostar, fielding2015}, and stellar flybys
\citep{batygin2020flyby, rodet2021sulai}.

The mechanism that is the focus of this paper was first proposed by
\citet{batygin2012} and involves a star, its surrounding PPD, and a distant
binary companion.
During the subsequent dissipation of the PPD, the secular precession frequency
of the stellar spin driven by disk and that of the disk driven by the binary
become commensurate, and the resulting secular resonance crossing excites the
obliquity to large values \citep{batygin2013adams, lai2014star,
spalding2014disk, zanazzi2018Paper1, zanazzi2018Paper2,
gerbig2024}.
The result of this mechanism is a nontrivial distribution of the primordial
obliquities.
The properties of this distribution have ramifications for the predictions of
subsequent secular evolution of the system, such as the von Zeipel-Lidov-Kozai
migration of giant planets \citep[e.g.][]{vick2023_HJ}. Thus, a detailed
characterization of the primordial obliquities is of great importance.

One attractive feature of this mechanism is that $\sim50\%$ of FGK stars are in
binaries \citep[e.g.][]{duquennoy1991, raghavan2010binary, tokovin2014}, with an
increase towards more massive stars \citep{offner2023review}.
Morever, binarity
statistics in young star-forming environments suggest that binaries may be more
common at early stellar ages \citep[e.g.
in the Taurus star-forming
region;][]{kraus2011taurus} and may be rarer in dense star clusters \citep[e.g.
in the Orion Nebula Cluster;][]{reipurth2007orion, duchene2018orion}.
Taken
together, these statistics suggest that the present-day FGK binary fraction of $50\%$
is likely to be an underestimate of their binarity at formation, which may
decrease to its present-day value due to dynamical scattering.
It is therefore
imperative to account for this disk dissipation-driven resonant obliquity
excitation mechanism, as it is likely to operate in a substantial fraction of
planet-hosting systems.

In this paper, we revisit this disk dissipation-driven misalignment scenario.
Our objectives are to provide a compact semi-analytical description of the
resulting obliquity distribution in the simplest approximation (i.e.\ a
homologously-disspating rigid disk), then to explore some variations of
this result under different initial conditions.
In particular, we study the effect of
non-homologous disk evolution, in which photoevaporation opens a gap at $\sim
2\;\mathrm{au}$.
We show that this non-homologous evolution changes the predicted obliquities for
planets with semimajor axes $\lesssim 1\;\mathrm{au}$ compared to the models
adopted in the literature.
In particular, while the standard disk models result in a prediction of general
misalignment for all but the closest Earth and Super Earth-mass planets
\citep[which is in disagreement with observation, e.g.][]{winn2017keplerobl,
albrecht2022review}, our model results in low primordial obliquities for warm
planets $\lesssim 1\;\mathrm{au}$ (weakly dependent on uncertain disk physics)
independent of planet mass.

In Section~\ref{s:unbroken_theory}, we review the essential theoretical results
for our analysis.
In Section~\ref{s:unbroken_dynamics}, we present a systematic exploration of the
obliquity evolution when assuming a standard, homologously-dissipating disk.
In Section~\ref{s:broken_disk}, we study the effect of including
photoevaporative gap opening in the disk evolution.
We summarize and discuss in Section~\ref{s:summary}.

\section{Theoretical Background: Precessional Dynamics of a Single, Rigid Disk
}\label{s:unbroken_theory}

We first discuss some relevant theoretical results when the PPD is treated as a
single, rigidly-precessing disk.
Our approach is modelled after that of \citet{zanazzi2018Paper1}.

\subsection{Equations of Motion}\label{ss:eom}

Our discussion here is expanded from what appears in \citet{vick2023_HJ}.
We consider a protostellar system consisting of a primary star $M_{\star}$
surrounded by a planet with mass $m_{\rm p}$ embedded in a dissipating
protoplanetary disk with total mass $M_{\rm d}$ and an external binary companion
with mass $M_{\rm b}$.
The star has radius $R_\star$, rotation rate $\Omega_{\rm \star}$, and spin
angular momentum vector $\bm{S} = S\uv{s}$, with
\begin{equation}
    S = k_\star M_\star R_\star^2 \Omega_\star,
\end{equation}
where the normalized moment of inertia $k_\star \simeq 0.2$ for a fully
convective star and $k_\star = 0.06$ for the Sun \citep{yoder1995}.
The star has a rotation-induced quadrupole moment $J_2 M_\star R_\star^2$ with
$J_2 = k_{\rm q\star} \Omega_\star^2 R_\star^3 / (GM\star)$.
We assume that the disk has a surface density profile given by
\begin{equation}
    \Sigma = \Sigma_{\rm in} \frac{r_{\rm in}}{r},
\end{equation}
which extends from $r_{\rm in}$ to $r_{\rm out}$.
Thus, the total disk mass is related to $\Sigma_{\rm in}$ by (assuming $r_{\rm
out} \gg r_{\rm in}$)
\begin{equation}
    M_{\rm d} \simeq 2 \uppi \Sigma_{\rm in} r_{\rm in} r_{\rm out}.
\end{equation}
The disk angular momentum vector is $\bm{L}_{\rm d} = L_{\rm d} \uv{l}_{\rm d}$
with
\begin{equation}
    L_{\rm d} \simeq \frac{2}{3} M_{\rm d}\sqrt{GM_\star r_{\rm out}}.
\end{equation}
The planet has a circular orbit with radius $a_{\rm p}$.
Throughout this paper, we assume that the planet's orbit axis $\uv{l}_{\rm p}$
is aligned with the disk axis $\uv{l}_{\rm d}$, i.e.\ $\uv{l}_{\rm p} =
\uv{l}_{\rm d} \equiv \uv{l}$; the reason for this assumption is fully discussed
in \citet{zanazzi2018Paper1}.
The binary companion $M_{\rm b}$ has an orbital radius $a_{\rm b}$ which is at
least a few times larger than $r_{\rm out}$.
Since $L_{\rm b} \gg L$ and $S$, we assume that $\uv{l}_{\rm b}$ remains fixed.

The system as described above has two dominant precessional effects: the mutual
precession of the star and the planet \& disk system, and the precession of the
planet \& disk system about the binary orbit (we ignore dissipation induced by
warps propagating within the disk).
The spin vector $\uv{s}$ evolves as
\begin{equation}
    \rd{\uv{s}}{t} = -\omega_{\rm sl}
        (\uv{l} \cdot \uv{s})(\uv{l} \times \uv{s}),
        \label{eq:dsdt}
\end{equation}
where $\omega_{\rm sl}$ is a combination of the spin-planet and spin-disk
precession frequencies:
\begin{align}
    \omega_{\rm sl} \equiv{}& \omega_{\rm sd} + \omega_{\rm sp}
        ,\label{eq:def_wsl}\\
    \omega_{\rm sp} ={}& \frac{3k_{\rm q\star}}{2k_\star}
            \p{\frac{m_{\rm p}}{M_\star}}
            \p{\frac{R_\star}{a_{\rm p}}}^3 \Omega_\star\nonumber\\
        ={}& \frac{2\pi}{1.8\;\mathrm{Gyr}}
            \p{\frac{2k_{\rm q\star}}{k_\star}}
            \p{\frac{m_{\rm p}}{M_{\rm J}}}
            \p{\frac{M_\star}{M_{\odot}}}^{-1}\nonumber\\
        &\times \p{\frac{R_\star}{2R_{\odot}}}^{3/2}
            \p{\frac{a_{\rm p}}{5\;\mathrm{au}}}^{-3}
            \p{\frac{P_\star}{3\;\mathrm{day}}}^{-1},\label{eq:def_wsp}\\
% 2 * pi / (yr * 3 / 4 * Mjup / Msun * ((2 Rsun) / (5 au))^3 * (2 * pi / (3 day)))
% 1.784611e+09
    \omega_{\rm sd} ={}& \frac{3k_{\rm q\star}}{2k_\star}
            \p{\frac{M_{\rm d}}{M_\star}}
            \p{\frac{R_\star^3}{r_{\rm in}^2r_{\rm out}}} \Omega_\star\nonumber\\
        ={}& \frac{2\pi}{10\;\mathrm{kyr}}
            \p{\frac{2k_{\rm q\star}}{k_\star}}
            \p{\frac{M_{\rm d}}{0.1M_\star}}
            \p{\frac{r_{\rm in}}{4R_{\star}}}^{-2}\nonumber\\
        &\times \p{\frac{r_{\rm out}}{50\;\mathrm{au}}}^{-1}
            \p{\frac{P_\star}{3 \;\mathrm{day}}}^{-1}
            \p{\frac{R_\star}{2R_{\odot}}}^{3/2},\label{eq:def_wsd}
% >>> 2 * pi / (yr * 3/8 * 0.1 * ((2 Rsun)^3 / ((8 Rsun)^2 * (50 au))) * (2 * pi / (3 day)))
% 1.884543e+04
\end{align}
where we've adopted a stellar radius that reflects the young age of the star and
a rotation rate consistent with those of young stars in the Orion Complex
\citep{kounkel2023youngspin}, and the disk properties are broadly consistent
with first-principles simulations of protostellar clump collapse
\citep{lebreuilly2024disk} and truncation at the stellar magnetosphere
\citep{lai2011mag, bouvier2020_magnotruncobs}.

The disk and planet experience gravitational torques from both the oblate star
and the binary companion.
We assume that the disk and planet remain strongly coupled at all times, as
their mutual gravitational precession is very rapid \citep{zanazzi2018Paper1}.
The combined disk and planetary axis $\uv{l}$ evolves according to
\begin{align}
    \rd{\uv{l}}{t} ={}&
        \omega_{\rm sl}\frac{S}{L}
            \p{\uv{l} \cdot \uv{s}} \p{\uv{l} \times \uv{s}}
        - \omega_{\rm lb} \p{\uv{l}_{\rm b} \cdot \uv{l}}
                            \p{\uv{l}_{\rm b} \times \uv{l}},\label{eq:dldt}
\end{align}
where $\omega_{\rm lb}$ is a combination of the planet-binary and disk-binary
precession:
\begin{align}
    \omega_{\rm lb} \equiv{}& \omega_{\rm db}\frac{L_{\rm d}}{L}
            + \omega_{\rm pb}\frac{L_{\rm p}}{L},\label{eq:def_wlb}\\
    \omega_{\rm db} ={}& \frac{3M_{\rm b}}{8M_\star}
            \p{\frac{r_{\rm out}}{a_{\rm b}}}^3 n_{\rm out}\nonumber\\
        ={}& \frac{2\pi}{0.2\;\mathrm{Myr}}
            \p{\frac{M_{\rm b}}{M_\star}}
            \p{\frac{M_\star}{M_{\odot}}}^{1/2}
            \p{\frac{r_{\rm out}}{50\;\mathrm{au}}}^{3/2}
            \p{\frac{a_{\rm b}}{300\;\mathrm{au}}}^{-3},\label{eq:def_wdb}\\
% 2 * pi / (yr * 3/8 * ((50 au) / (300 au))^3 * (G * (Msun) / (50 au)^3)^(1/2))
% 2.036349e+05
    \omega_{\rm pb} ={}& \frac{3M_{\rm b}}{4M_\star}
            \p{\frac{a_{\rm p}}{a_{\rm b}}}^3 n_{\rm p}\nonumber\\
        ={}& \frac{2\pi}{3.2\;\mathrm{Myr}}
            \p{\frac{M_{\rm b}}{M_\star}}
            \p{\frac{M_\star}{M_{\odot}}}^{1/2}
            \p{\frac{a_{\rm p}}{5\;\mathrm{au}}}^{3/2}
            \p{\frac{a_{\rm b}}{300\;\mathrm{au}}}^{-3},\label{eq:def_wpb}
% 2 * pi / (yr * 3/4 * ((5 au) / (300 au))^3 * (G * (Msun) / (5 au)^3)^(1/2))
% 3.219750e+06
\end{align}
and $L \equiv L_{\rm p} + L_{\rm d}$ is the total angular momentum of the
combined disk and planet. The angular momentum ratios are:
\begin{align}
    \frac{S}{L_{\rm p}} ={}& \frac{k_\star M_\star R_\star^2 \Omega_\star}{
            m_{\rm p} \sqrt{GM_\star a_{\rm p}}}\nonumber\\
% >>> 0.1 * Msun * (2 Rsun)^2 * (2 * pi / (3 day)) / (Mjup * (G * Msun * (5 au))^(1/2))
% 0.493261
        ={}& 0.5
            \p{\frac{M_\star}{M_{\odot}}}^{1/2}
            \p{\frac{k_\star}{0.1}}
            \p{\frac{R_\star}{2 R_{\odot}}}^2
            \p{\frac{P_\star}{3\;\mathrm{day}}}^{-1}\nonumber\\
        &\times \p{\frac{m_{\rm p}}{M_{\rm J}}}^{-1}
            \p{\frac{a_{\rm p}}{5\;\mathrm{au}}}^{-1/2},\\
    \frac{S}{L_{\rm d}} ={}& \frac{k_\star M_\star R_\star^2 \Omega_\star}{(2/3)
            M_{\rm d} \sqrt{GM_\star r_{\rm out}}}\nonumber\\
% 0.1 * (2 Rsun)^2 / (0.1 * 2 / 3 * (G * Msun * (50 au))^(1/2)) * (2 * pi / (3 day))
% 0.002233
        ={}& \scinot{2.2}{-3}
            \p{\frac{M_\star}{M_{\odot}}}^{1/2}
            \p{\frac{k_\star}{0.1}}
            \p{\frac{R_\star}{2 R_{\odot}}}^2
            \p{\frac{P_\star}{3\;\mathrm{day}}}^{-1}\nonumber\\
        &\times \p{\frac{M_{\rm d}}{0.1M_{\star}}}^{-1}
            \p{\frac{r_{\rm out}}{50\;\mathrm{au}}}^{-1/2}.
\end{align}
For the fiducial parameters, the breakup rotation rate of the star is
$\sqrt{GM_\star / R_\star^3} = 2\pi / \p{0.33\;\mathrm{days}}$.
Eq.~\eqref{eq:def_wlb} differs slightly from Eq.~(69) of
\citet{zanazzi2018Paper1} (which assumes $L_{\rm d} \gg L_{\rm p}$) and
describes the combined precession of a planet-disk system slightly more
accurately.

For convenience, we define the three mutual misalignment angles
\begin{align}
    \cos \theta_{\rm lb} &\equiv \uv{l}_{\rm b} \cdot \uv{l}&
    \cos \theta_{\rm sl} &\equiv \uv{s} \cdot \uv{l}&
    \cos \theta_{\rm sb} &\equiv \uv{s} \cdot \uv{l}_{\rm b}.
    \label{eq:def_I}
\end{align}
Together, these three angles completely define the relative orientations of the
three vectors.

\subsection{$S \ll L$ Equilibria: Colombo's Top and Cassini
States}\label{ss:th_sllL}

If $S \ll L$ at all times, then the system as described by Eqs.~\eqref{eq:dsdt}
and~\eqref{eq:dldt} reduces to the well-studied ``Colombo's Top'' system
\citep{colombo1966, peale1969, peale1974possible, henrard1987, ward2004I,
su2020_disk, su2022_weaktide, su2022_multics}. We briefly recap the properties of this
system and our notations below; see \citet{su2020_disk} for a more thorough
discussion.

Colombo's Top admits solutions where $\uv{s}$ and $\uv{l}$ precess in a way
such that $\uv{s}$, $\uv{l}$, and $\uv{l}_{\rm b}$ remain coplanar. When
the system satisfies this condition, $\uv{s}$ is said to be in one of the
``Cassini States'' (CSs). We follow \citet{su2020_disk} and define the
dimensionless parameter
\begin{equation}
    \eta \equiv \frac{\omega_{\rm lb}}{\omega_{\rm sl}} \cos \theta_{\rm lb}.
        \label{eq:def_eta}
\end{equation}
Note that $\theta_{\rm lb}$ is constant when $S \ll L$ (Eq.~\ref{eq:dldt}).
There are two CSs for $\eta > \eta_{\rm c}$ and four for $\eta < \eta_{\rm c}$
where \citep{ward2004I}
\begin{equation}
    \eta_{\rm c} = \p{\sin^{2/3} \theta_{\rm lb} + \cos^{2/3}\theta_{\rm lb}}^{3/2}.
        \label{eq:def_etac0}
\end{equation}

\begin{figure}
    \centering
    \includegraphics[width=\columnwidth]{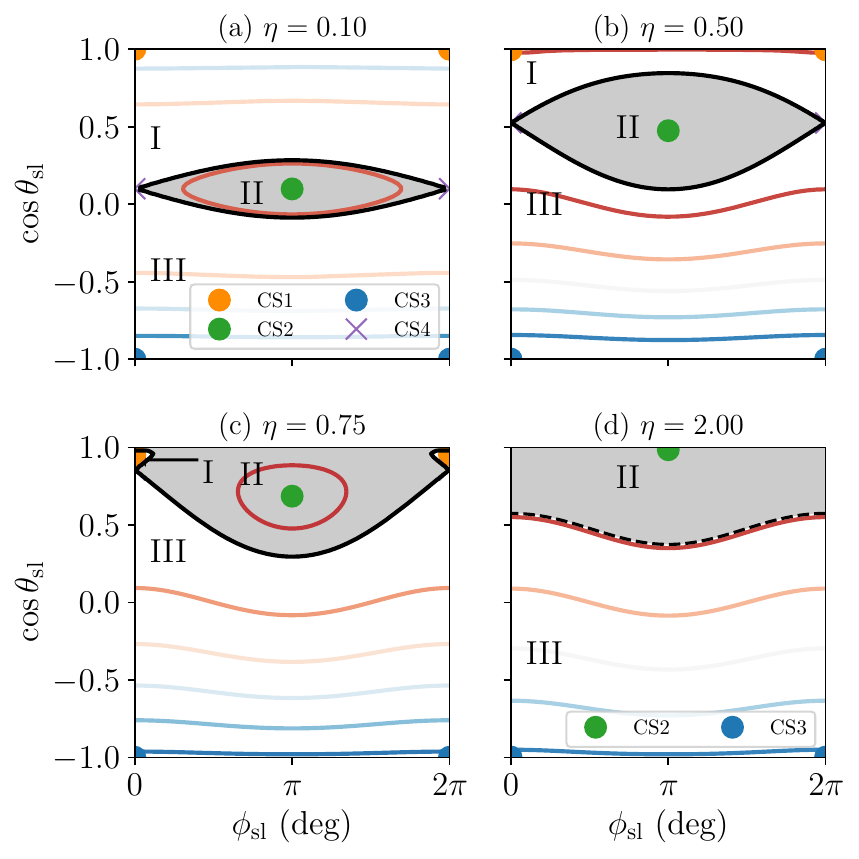}
    \caption{
    Phase portrait of the stellar spin dynamics when $S / L \ll 1$ for a few
    values of $\eta$ (Eq.~\ref{eq:def_eta}).
    Here, we have taken $\theta_{\rm lb} = 5^\circ$, for which $\eta_{\rm c}
    \approx 0.77$.
    The contours depict level curves of the conserved Hamiltonian given by
    Eq.~\eqref{eq:H}, which are also the trajectories along which the spin
    evolves.
    Note that we have labelled as colored dots the four Cassini States for $\eta
    < \eta_{\rm c}$ and the two for $\eta > \eta_{\rm c}$ (lower right panel).
    In these first three panels, the boundary of the shaded region is the
    \emph{separatrix}, which divides phase space into the three labeled regions
    I/II/III\@.
    In the lower right panel, there is no separatrix, but the dashed line
    encloses the same area as the separatrix encloses when $\eta = \eta_{\rm c}$
    (given by Eq.~\ref{eq:asep_S0}).
    }\label{fig:contours}
\end{figure}

To describe the phase space over which $\uv{s}$ evolves, we transform into the
non-inertial reference frame corotating with $\uv{l}$ about $\uv{l}_{\rm b}$
such that both $\uv{l}$ and $\uv{l}_{\rm b}$ are fixed.
We then adopt the spherical coordinate system where $\uv{z}$ axis is aligned
with $\uv{l}$, the polar angle is $\theta_{\rm sl} \in \s{0, 180^\circ}$, and
the azimuthal angle $\phi_{\rm sl}$ is defined relative to the
$\uv{l}$-$\uv{l}_{\rm b}$ plane such that $\phi_{\rm sl} = 0$ corresponds to
$\uv{s}$ and $\uv{l}_{\rm b}$ being on opposite sides of $\uv{l}$.
Note that $\p{\cos \theta_{\rm sl}, \phi_{\rm sl}}$ form a canonically conjugate
pair of variables, and the equations of motion can be derived from the
Hamiltonian
\begin{align}
    H
        ={}& -\frac{\omega_{\rm sl}}{2} \cos^2\theta_{\rm sl}
                + \omega_{\rm lb}\cos \theta_{\rm lb}\nonumber\\
            &\times
                \p{\cos\theta_{\rm sl} \cos \theta_{\rm lb}
                - \sin \theta_{\rm lb} \sin\theta_{\rm sl} \cos \phi_{\rm sl}}.
                \label{eq:H}
\end{align}
The phase portrait for the Colombo's Top dynamics in these coordinates is shown
in Fig.~\ref{fig:contours}.

In these coordinates, $\theta_{\rm sl}$ for every CS satisfies the following
relation \citep{peale1969, ward1975tidal}
\begin{equation}
    % \sin \theta_{\rm sl} \cos \theta_{\rm sl}
    %     - \eta\p{
    %         \cos \theta_{\rm lb} \sin \theta_{\rm sl}
    %         + \sin \theta_{\rm lb} \cos \theta_{\rm sl} \cos \phi_{\rm sl}
    %     }
    %     = 0.\label{eq:CS2_cond}
    \sin \theta_{\rm sl} \cos \theta_{\rm sl}
        - \eta \sin\p{\theta_{\rm sl} \pm \theta_{\rm lb}} = 0.\label{eq:CS2_cond}
\end{equation}
In this expression, the positive sign is taken for $\phi_{\rm sl} = 0$ (CSs 1,
3, 4), and the negative sign is taken for $\phi_{\rm sl} = \pi$ (CS2).

In Fig.~\ref{fig:contours}, the panels where $\eta \leq \eta_{\rm c}$ contain a
\emph{separatrix}---the boundary across which $\phi_{\rm sl}$ changes from
librating to circulating---denoted as the thick black line. While the phase
space area enclosed by the separatrix can be expressed in closed form for all
$\eta \leq \eta_{\rm c}$ \citep{ward2004I}, it simplifies for $\eta = \eta_{\rm
c}$ to:
\begin{align}
    A_{\rm sep, c} = 4\pi\s{1
        - \p{1 + \tan^{2/3}\theta_{\rm lb}}^{-3/2}}.\label{eq:asep_S0}
\end{align}
When a system is initially in CS1 and $\eta$ increases to be $> \eta_{\rm c}$,
the system will transition to a trajectory enclosing phase space area $A_{\rm
sep, c}$. An example of such a trajectory is depicted by the black dashed line
(shaded) in the lower-right panel of Fig.~\ref{fig:contours}.

\subsection{Equilibria for Comparable $S$ and $L$}\label{ss:th_ssimL}

\begin{figure}
    \centering
    \includegraphics[width=\columnwidth]{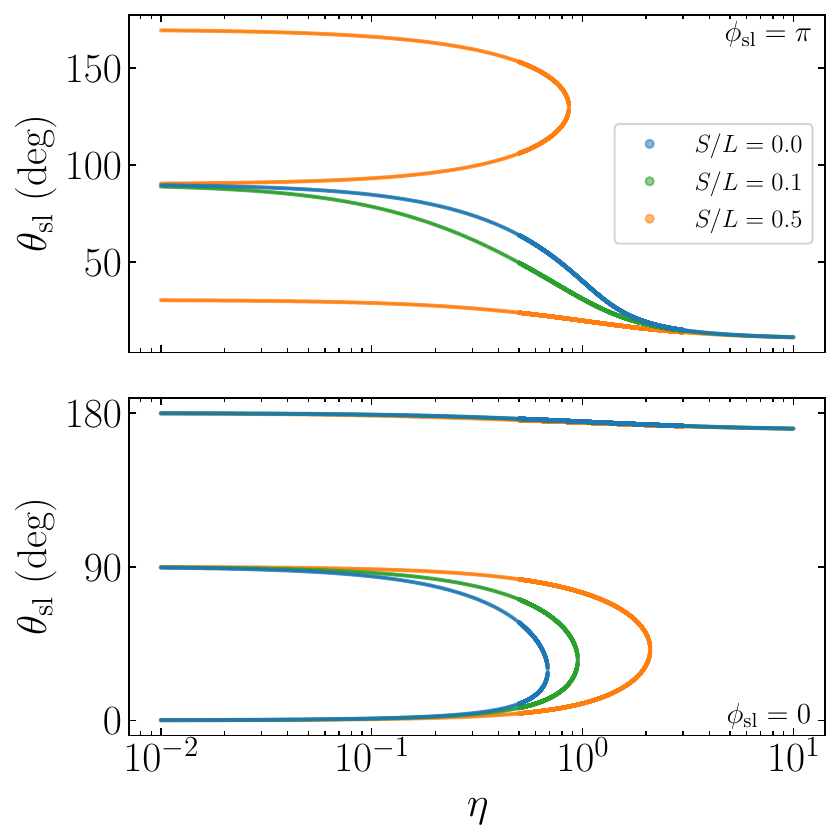}
    \caption{
    Cassini State-like equilibria for $\theta_{\rm lb} = 10^\circ$, obtained by
    solving Eq.~\ref{eq:gCS2_cond} for the labeled values of $S / L$.
    The top panel shows the equilibria for $\phi_{\rm sl} = \pi$ (including CS2
    for $S = L = 0$) and the bottom panel shows the equilibria for $\phi_{\rm
    sl} = 0$ (including CSs 1, 3, and 4 for $S / L = 0$).
    With increasing $S / L$, the classical four CSs are slightly modified while
    new equilibria appear.
    }\label{fig:css}
\end{figure}

The case when $S / L$ is non-negligible (but both $S$ and $L$ are still $\ll
L_{\rm b}$) has been studied less than Colombo's Top, though excellent works
have discussed this regime \citep[e.g.][]{goldreich1966history,
correia2015stellar}.
We follow the treatment given in \citet{anderson2018teeter}, where we combine
Eqs.~\eqref{eq:dsdt} and~\eqref{eq:dldt} and obtain a general condition for
CS-like equilibria
\begin{equation}
    \rd{}{t}\s{\uv{s} \cdot \p{\uv{l} \times \uv{l}_{\rm b}}} = 0.
\end{equation}
This can be reduced to the algebraic form
\begin{align}
    \cos&\theta_{\rm sl} \sin\theta_{\rm sl} \sin \theta_{\rm lb}\nonumber\\
        &- \p{\frac{S}{L}\cos\theta_{\rm sl}\sin\theta_{\rm sl}
            + \eta \sin \theta_{\rm lb}}\sin\p{\theta_{\rm sl} \pm \theta_{\rm
            lb}} = 0.\label{eq:gCS2_cond}
\end{align}
Here, again, the positive sign corresponds to $\phi_{\rm sl} = 0$.
Note that
this reduces to Eq.~\eqref{eq:CS2_cond} for $S = 0$.
This equation can be
obtained by evaluating Equation~(19) of \citet{anderson2018teeter} when their
$L_1 \ll L_{\rm p}$ ($L \ll L_{\rm b}$ in our notation).
The locations of all
equilibria for a few characteristic $S / L$ are shown in
Fig.~\ref{fig:css} \citep[c.f.
Fig~6 of][]{anderson2018teeter}.

\section{Single Disk: Dynamics During Disk Dissipation}\label{s:unbroken_dynamics}

In the previous section, we have reviewed the precessional dynamics of the
star-planet-disk-binary system when the precession frequencies are fixed.
In this section, we will study the evolution of the system as the disk
dissipates.

\subsection{Obliquity Excitation}

We assume that the disk dissipates homologously, with its total mass evolving
as:
\begin{equation}
    M_{\rm d}(t) = 0.1M_\star e^{-t / \tau_{\rm d}}.\label{eq:md}
\end{equation}
A typical value for the disk e-folding time is $\tau_{\rm d} \sim \mathrm{Myr}$
\citep{bertout2007_disklifetime, galli2015_disklifetime}.
Eq.~\eqref{eq:md}, in conjunction with Eqs.~(\ref{eq:dsdt}) and~(\ref{eq:dldt}),
fully describe the evolution of the combined star-planet-disk-binary system as
the disk dissipates.

\begin{figure}
    \centering
    \includegraphics[width=\columnwidth]{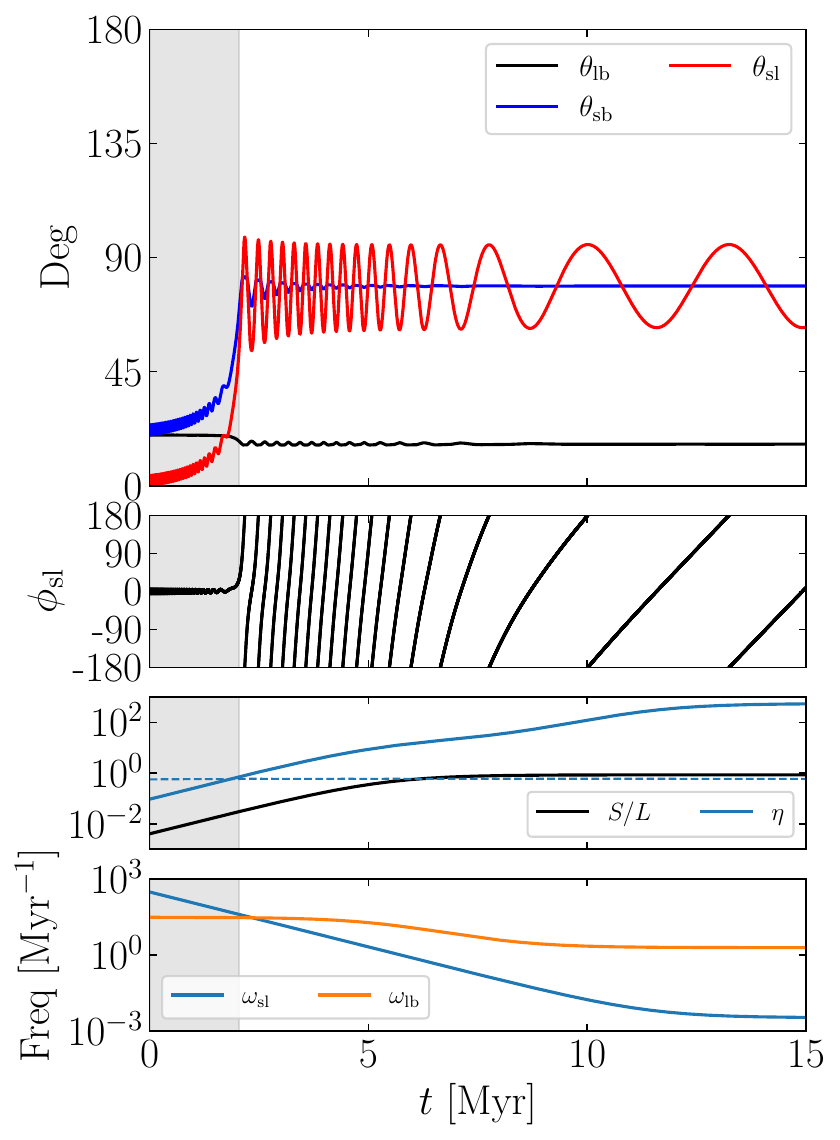}
    \caption{
    Example of the obliquity excitation process obtained by integrating
    Eqs.~\eqref{eq:dsdt},~\eqref{eq:dldt}, and~\eqref{eq:md}.
    In the first panel, we show the evolution of the mutual misalignment angles
    (Eq.~\ref{eq:def_I}).
    Obliquity excitation can be seen in the growth of $\theta_{\rm sl}$ (red)
    from near zero to large values by the end of the integration.
    In the second panel, we show the angle $\phi_{\rm sl}$, which librates when
    the system is in a CS (librating times are shaded grey).
    In the third panel, we show $S / L$, the ratio of the stellar spin angular
    momentum to the combined disk-planet orbital angular momentum, and $\eta$
    (Eq.~\ref{eq:def_eta}).
    The value of $\eta_{\rm c}$ (Eq.~\ref{eq:def_etac0}) is shown as the blue
    dashed line.
    In the last panel, we show the evolution of the precession frequencies
    $\omega_{\rm sl}$ (Eq.~\ref{eq:def_wsl}) and $\omega_{\rm lb}$
    (Eq.~\ref{eq:def_wlb}).
    }\label{fig:sample}
\end{figure}

In Fig.~\ref{fig:sample}, we show an example of how a system evolves in terms of
the misalignment angles of the system (Eq.~\ref{eq:def_I}). Obliquity excitation
can be seen by the growth of $\theta_{\rm sl}$ (red) from zero to large values.
The second panel depicts the initial libration of $\phi_{\rm sl}$ until resonant
obliquity excitation; libration is denoted in all panels by the grey shaded
regions. This corresponds with the moment when $\eta$ increases beyond
$\eta_{\rm c}$ (third panel); note that $\eta_{\rm c}$ is not exactly constant
due to the slight evolution of $\theta_{\rm lb}$.

\subsection{Solution for $S \ll L$}\label{ss:sol_sllL}

To begin understanding the dynamics of the system, we first consider the limit
where $S \ll L$, and the dynamics reduces to Colombo's Top
(Section~\ref{ss:th_sllL}).
We describe the evolution of the system in terms of
the two misalignment angles $\theta_{\rm sb}$ and $\theta_{\rm sl}$; note
that $\theta_{\rm lb}$ is constant when $S \ll L$ (Eq.~\ref{eq:dldt}).
For our analytical results, we further assume that the disk dissipation is
adiabatic---i.e.\ that $\tau_{\rm d}$ is much longer than all precessional
timescales in the system.
For initial conditions, we take the initial stellar spin and disk axes to be
aligned such that the initial $\theta_{\rm sl, i} = 0$.
This is the expected configuration at star formation, though there may be
additional physical mechanisms that induce early spin-disk misalignment (e.g.\
accretion misalignment, magnetic warping; see \citealp{albrecht2022review} for a
review).

Then, as the disk dissipates, $\eta$ increases from its initial value
% >>> 3 / 8 * (50 / 300)^3 * (G * Msun / (50 au)^3)^(1/2)/ (3/8 * 0.1 * (2 / 8)^2 * (2 Rsun) / (50 au) * 2 * pi) * (3 day)
% 0.092545
\begin{align}
    \eta_{\rm i} \approx {}& \frac{\omega_{\rm db}}{\omega_{\rm sd}}
            \cos\theta_{\rm lb}\nonumber\\
        ={}& 0.09
            \p{\frac{k_\star}{4k_{\rm q\star}}}
            \p{\frac{r_{\rm out}}{50\;\mathrm{au}}}^{5/2}
            \p{\frac{r_{\rm in}}{4R_\star}}^2
            \p{\frac{R_\star}{2R_\odot}}^{-3}
            \nonumber\\
        &\times \p{\frac{a_{\rm b}}{300\;\mathrm{au}}}^{-3}
            \p{\frac{M_{\rm d}}{0.1M_{\odot}}}^{-1}
            \p{\frac{P_\star}{3 \;\mathrm{day}}}
            \cos \theta_{\rm lb},
            \label{eq:eta_i}
\end{align}
to its final value
% >>> (3 / 4 * (5 / 300)^3 * (G * Msun / (5 au)^3)^(1/2)) / (3/8 * (Mjup / Msun) * ((2 Rsun) / (5 au))^3 * 2 * pi / (3 day))
% 1108.540202
\begin{align}
    \eta_{\rm f} \approx {}& \frac{\omega_{\rm pb}}{\omega_{\rm sp}}
            \cos \theta_{\rm lb}\nonumber\\
        ={}& 1000
            \p{\frac{k_\star}{4k_{\rm q\star}}}
            \p{\frac{M_\star}{M_{\odot}}}^{1/2}
            \p{\frac{m_{\rm p}}{M_{\rm J}}}^{-1}
            \p{\frac{a_{\rm p}}{5 \;\mathrm{au}}}^{9/2}\nonumber\\
        &\times \p{\frac{a_{\rm b}}{300\;\mathrm{au}}}^{-3}
            \p{\frac{2R_\star}{R_{\odot}}}^{-3}
            \p{\frac{P_\star}{3 \;\mathrm{day}}}
            \cos \theta_{\rm lb}.
            \label{eq:eta_f}
\end{align}
We have assumed an equal-mass binary companion $M_{\rm b} = M_\star$.

\begin{figure}
    \centering
    \includegraphics[width=\columnwidth]{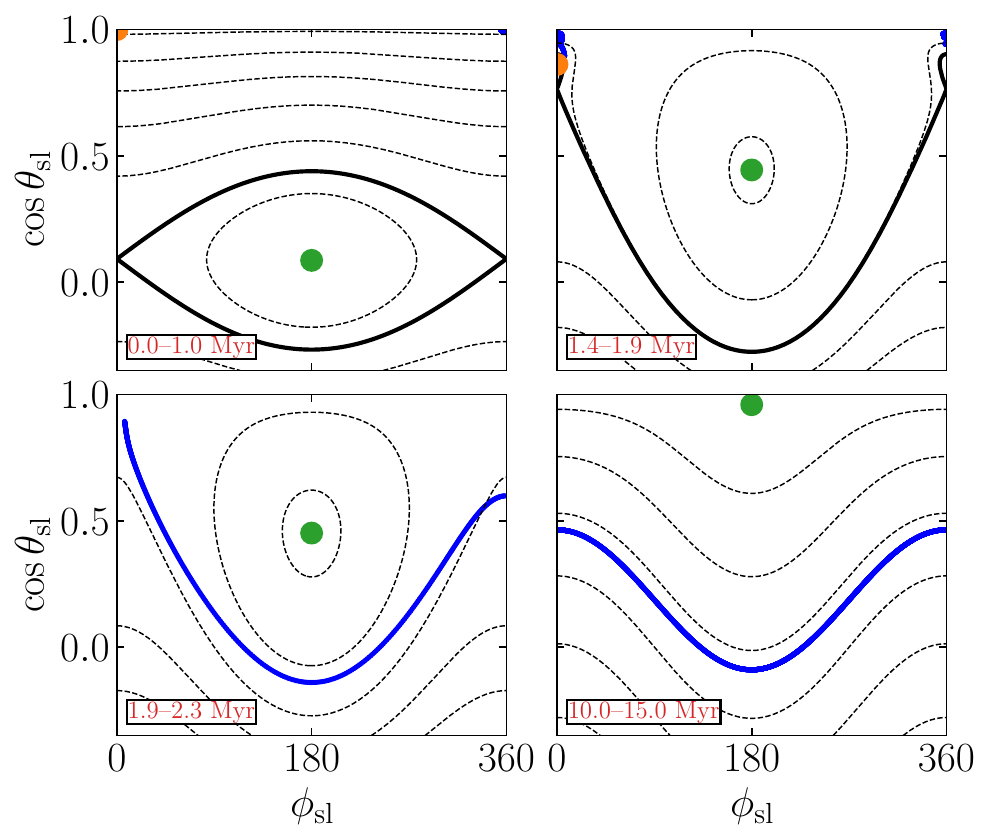}
    \caption{
    Phase space evolution corresponding to Fig.~\ref{fig:sample}.
    The blue lines denote the evolution of the stellar spin axis over the
    interval of time denoted in the bottom left of each panel.
    Note the ejection from CS1 (orange) between the second and third panels, and
    the large-amplitude precession about CS2 (green) at late times in the fourth
    panel.
    The black dashed lines represent level curves of the Colombo's Top
    Hamiltonian (Eq.~\eqref{eq:H}), and the black solid line the separatrix
    (when it exists).
    The slight discrepancies between the numerical integrations and level curves
    are due to the ongoing evolution of the system as well as the nonzero
    angular momentum ratio $S / L$.
    }\label{fig:sample_phase}
\end{figure}

The basic features of the fiducial evolution in Fig.~\ref{fig:sample} can be
understood by a simple analysis of the phase space evolution, see
Fig.~\ref{fig:contours}:
\begin{itemize}
    \item Since $\theta_{\rm sl, i} = 0$, the system occupies CS1 initially, and
        $\phi_{\rm sl}$ librates.

    \item Then, as $\eta$ increases, the system remains in CS1 until $\eta =
        \eta_{\rm c}$, at which point CS1 and CS4 disappear via a saddle-node
        bifurcation, and the system exits the CS\@; this also corresponds to the
        onset of circulation in $\phi_{\rm sl}$.
        From this point forward, $\uv{s}$ begins to precess about CS2, enclosing
        the phase space area given by Eq.~\eqref{eq:asep_S0}.

        Since the dynamics are symmetric under the reflection $\uv{l}_{\rm b}
        \to -\uv{l}_{\rm b}$, Eq.~\eqref{eq:asep_S0} is also accurate for
        $\theta_{\rm lb} > 90^\circ$.

    \item As $\eta$ continues to increase, the enclosed phase space area is an
        adiabatic invariant.
        For large $\eta$, CS2 is aligned with $\uv{l}_{\rm b}$
        (Fig.~\ref{fig:css}), and the spin simply precesses about it with
        constant $\theta_{\rm sb} = \theta_{\rm sb, f}$.
\end{itemize}
This phase space evolution is illustrated in
Fig.~\ref{fig:sample_phase}, where the ejection from the stable CS1 and the
resulting large-amplitude precession about CS2 is evident
\citep[cf.\ Fig.~4 of][]{anderson2018teeter}.

Equating the enclosed phase
space area at this time with Eq.~\eqref{eq:asep_S0}, we obtain
[cf.\ Eq.~(16) of \citealp{ward2004I}],
\begin{align}
    \cos \theta_{\rm sb, f} =
    \begin{cases}
        \frac{2}{\p{1 + \tan^{2/3} \theta_{\rm lb}}^{3/2}} - 1
            & \theta_{\rm lb} < 90^\circ,\\
        1 - \frac{2}{\p{1 + \tan^{2/3}
            (180^\circ - \theta_{\rm lb})}^{3/2}}
            & \theta_{\rm lb} > 90^\circ.
    \end{cases}\label{eq:qsbf_S0}
\end{align}

Since $\theta_{\rm lb}$ is constant in the $S \ll L$ limit, we can also solve
for the range of $\theta_{\rm sl, f}$ by simple geometry: it can vary within the
bounds \citep{spalding2014disk, anderson2018teeter}
\begin{align}
    \theta_{\rm sl, f} &\geq  \abs{\theta_{\rm lb} - \theta_{\rm sb, f}}
        \nonumber\\
    \theta_{\rm sl, f} &\leq \min\p{\theta_{\rm lb} + \theta_{\rm sb, f},
        360^\circ - \p{\theta_{\rm lb} + \theta_{\rm sb, f}}}
            \label{eq:qslf_triangle}.
\end{align}
In fact, $\theta_{\rm sl, f}$ spans this full range as $\uv{s}$ precesses about
$\uv{l}_{\rm b}$.

In Fig.~\ref{fig:tau3plots}, we compare the distributions obtained via
Eqs.~\eqref{eq:qsbf_S0} and~\eqref{eq:qslf_triangle} [black dashed lines] to the
values obtained via full numerical integrations of Eqs.~\eqref{eq:dsdt}
and~\eqref{eq:dldt} evaluated at $t = 10\tau_{\rm d} = 10\;\mathrm{Myr}$ (red
lines) for $2000$ values of $\theta_{\rm lb}$.
Satisfactory agreement is obtained for much of the parameter space, but notable
disagreement is obtained when $\theta_{\rm lb, i}$ is near $0^\circ$,
$90^\circ$, or $180^\circ$.
Nevertheless, the key features of the dynamics (e.g.\ the broadly retrograde
distribution of $\theta_{\rm sl, f}$ in the bottom panel of
Fig.~\ref{fig:tau3plots}) are reproduced.
In the next subsection, we expand the model slightly and obtain improved
agreement for all $\theta_{\rm lb}$.

\subsection{Comparable $S$ and $L$: Angular Momentum Budget}\label{ss:sol_ssimL}

The results of the previous subsection is not exact: as the disk dissipates, the
total $L$ decreases, and the system goes from the disk-dominated regime where $S
\ll L \approx L_{\rm d}$ to the planet-dominated regime where $S \sim L \approx
L_{\rm p}$ (see values in Section~\ref{ss:eom}).
In this case, $\theta_{\rm lb}$ is no longer constant as the disk dissipates,
and the final configuration of the system is defined by the two angles
$\theta_{\rm lb}$ and $\theta_{\rm sb}$.
To be precise, we wish to understand the final values $\theta_{\rm lb, f}$ and
$\theta_{\rm sb, f}$ as a function of the initial values $\theta_{\rm lb, i}$
and $\theta_{\rm sl, i}$ ($\theta_{\rm sb}$ is rapidly varying at early times,
while $\theta_{\rm sl}$ is rapidly varying at late times).

The primary difficulty in understanding these dynamics lies in the coupled
evolution of $S / L$ and $\theta_{\rm lb}$.
To circumvent this difficulty, we propose the following approximation.
Note that the phase space structure is largely insensitive to the specific value
of $\eta$ if either $\eta \gg \eta_{\rm c}$ or $\eta \ll \eta_{\rm c}$, while it
is sensitively dependent on $\eta$ when $\eta \sim \eta_{\rm c}$.
Motivated by this observation, we approximate that $\theta_{\rm
lb}$ and $\theta_{\rm sl}$ are approximately constant when $\eta \gtrsim
\eta_{\rm c}$, while $\theta_{\rm lb}$ and $\theta_{\rm sb}$ are approximately
constant when $\eta \lesssim \eta_{\rm c}$; in the intermediate short interval
where $\eta \sim \eta_{\rm c}$, we approximate that the variation in $L$ can be
neglected.
Under this approximation, the conservation of angular momentum along
$\uv{l}_{\rm b}$ during the $\eta \sim \eta_{\rm c}$ interval
(where the phase space bifurcation occurs) requires\footnote{
When the time interval of consideration is sufficiently short
that the change to $L$ can be neglected,
the angular momentum of the $\bm{S} + \bm{L}$ system along
$\uv{l}_{\rm b}$ is conserved because the torque due to $\uv{l}_{\rm b}$ is a
purely precessional one (Eq.~\ref{eq:dldt}): it does not act on the aligned
components of either $\bm{S}$ or $\bm{L}$.
} that the final and initial spin orientations be related as:
\begin{equation}
    S_{\rm c}\cos \theta_{\rm sb, i} + L_{\rm c} \cos \theta_{\rm lb, i}
    \approx
    S_{\rm c}\cos \theta_{\rm sb, f} + L_{\rm c} \cos \theta_{\rm lb, f}.
    \label{eq:lcons_etac}
\end{equation}
Here, the c subscript denotes that $S$ and $L$ are to be evaluated where $\eta =
\eta_{\rm c}$.
Under the assumption of initial spin-disk alignment ($\theta_{\rm sb, i} =
\theta_{\rm lb, i}$), we can simplify Eq.~\eqref{eq:lcons_etac} to obtain
\begin{equation}
    \cos \theta_{\rm lb, f}{} -{} \cos \theta_{\rm lb, i}
        = \p{\frac{S}{L}}_{\rm c}
            \s{\cos \theta_{\rm lb, i}
            - \cos \theta_{\rm sb, f}},\label{eq:final_constr1}
\end{equation}
where for brevity we've denoted $(S / L)_{\rm c} \equiv S_{\rm c} /
L_{\rm c}$, the angular momentum ratio evaluated at the location of the
bifurcation where $\eta = \eta_{\rm c}$.
To solve Eq.~\eqref{eq:final_constr1}, it is necessary to know the disk mass
$M_{\rm d}$ (and $L$) at which $\eta = \eta_{\rm c}$.
In general, $\eta_{\rm c}$ also depends on $S / L$ (see Fig.~\ref{fig:css}), but
for simplicity we adopt the value of $\eta_{\rm c}$ at $S = 0$
(Eq.~\ref{eq:def_eta}), which is sufficiently accurate for our analysis.
Now, if $(S / L)_{\rm c} \ll 1$, we can solve
Eq.~\eqref{eq:final_constr1} perturbatively: we can approximate $\theta_{\rm sb,
f}$ using Eq.~\eqref{eq:qsbf_S0} (which is exact for $S = 0$), then use
Eq.~\eqref{eq:final_constr1} to obtain the corresponding $\theta_{\rm lb, f}$.

\begin{figure}
    \centering
    \includegraphics[width=\columnwidth]{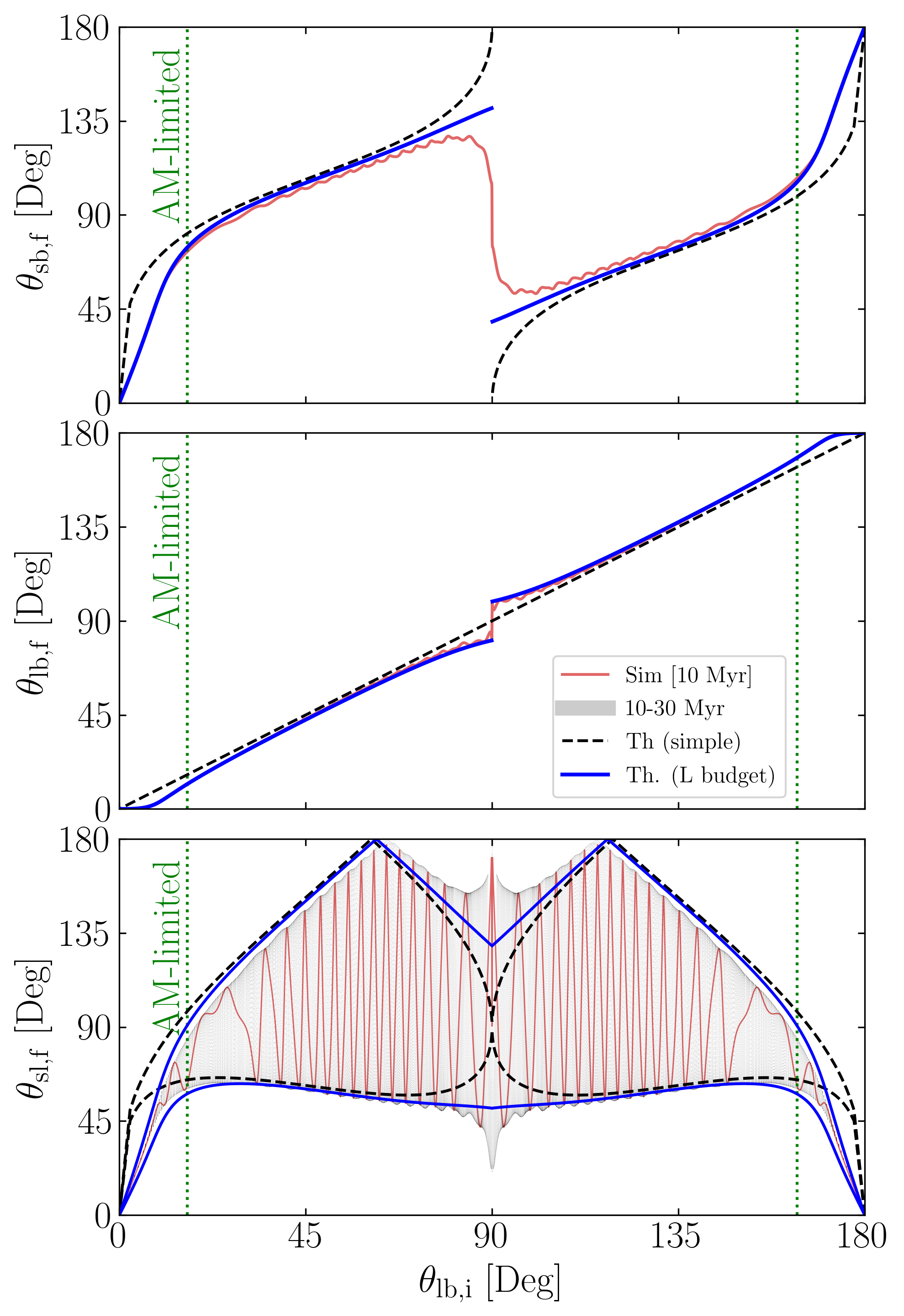}
    \caption{
    The final misalignment angles $\theta_{\rm sb, f}$, $\theta_{\rm lb, f}$,
    and $\theta_{\rm sl, f}$ (Eqs.~\ref{eq:def_I}) after the protoplanetary disk
    has dissipated (evaluated at $10\;\mathrm{Myr}$) for $2000$ values of the
    initial disk-binary misalignment angle $\theta_{\rm lb, i}$.
    The spin and disk are initially aligned ($\theta_{\rm sl, i} = 0$).
    We use a disk dissipation timescale of $\tau_{\rm d} = 1\;\mathrm{Myr}$.
    In all three panels, the black dashed line represents the $S \ll L$
    analytical expressions given by Eqs.~\eqref{eq:qsbf_S0}
    and~\eqref{eq:qslf_triangle} at fixed $\theta_{\rm lb}$,
    while the more accurate blue solid line is obtained by
    simultaneous solution of Eqs.~\eqref{eq:qsbf_S0},~\eqref{eq:qslf_triangle},
    and~\eqref{eq:final_constr1} (as discussed in Section~\ref{ss:sol_ssimL}).
    These two expressions primarily differ when the disk does not have
    sufficient angular momentum to tilt the stellar spin, notated by the green
    vertical dashed lines in all panels (Eq.~\ref{eq:Ii_near0}).
    In the bottom panel, the grey shaded region denotes the range of
    oscillation of the obliquity over $10$--$30\;\mathrm{Myr}$.
    Note that the small fluctuations in all three panels for nearby values of
    $\theta_{\rm lb, i}$ are due to the nonadiabatic effect of a finite
    $\tau_{\rm d}$.
    }\label{fig:tau3plots}
\end{figure}

The improved accuracy of this treatment to the $S \ll L$ solution presented in
Section~\ref{ss:th_sllL} can be seen in the solid blue lines in
Fig.~\ref{fig:tau3plots}.
A noticeable improvement is obtained, and the largest remaining inaccuracy lies
in the values of $\theta_{\rm sb, f}$ near $\theta_{\rm lb, i} \approx
90^\circ$.
This range of angles corresponds to the regime where the $S \simeq L$
corrections are largest, and our approximate perturbative treatment is no longer
accurate.

While Eq.~\eqref{eq:final_constr1} has no general closed-form solution, it can
be solved in a few limiting cases. We restrict our attention to $\theta_{\rm lb,
i} < 90^\circ$ (the symmetry of the problem yields straightforward extensions
to all $\theta_{\rm lb, i}$ values) and consider the following cases:

\textbf{(i)}
When $\theta_{\rm lb, i} \sim 90^\circ$, then $\cos \theta_{\rm sb, f}
\approx -1$,
and we obtain
\begin{equation}
    (S / L)_{\rm c} + \cos \theta_{\rm lb, i} = \cos \theta_{\rm lb, f}.
        \label{eq:angmom_budget_90}
\end{equation}
However, when $\cos \theta_{\rm lb}$ is small, the orbital precession is much
slower, and the disk must dissipate for a longer time before $\eta = \eta_{\rm
c}$ is reached; this shows that $(S / L)_{\rm c}$ depends on the angle
$\theta_{\rm lb}$.
To proceed, it is easiest to define
\begin{align}
    \p{\frac{S}{L}}_{\rm c0}
        &\equiv
    \p{\frac{S}{L}}_{\rm c}\cos \theta_{\rm lb, c}\nonumber\\
        &\approx \frac{1}{24}
            \p{\frac{M_\star}{M_{\rm b}}}
            \p{\frac{r_{\rm out}}{50\;\mathrm{au}}}^{-3}
            \p{\frac{r_{\rm in}}{8R_\odot}}^{-2}
            \p{\frac{a_{\rm b}}{300\;\mathrm{au}}}^{3},
            \label{eq:def_slc}
\end{align}
i.e.\ $(S / L)_{\rm c0}$ is the value of $S / L$ evaluated for $\eta = \eta_{\rm
c}$ and $\theta_{\rm lb, c} = 0$; note that we have only included the disk and
binary dependencies in Eq.~\eqref{eq:def_slc} for brevity.
% TODO here
Then, if we approximate that $\theta_{\rm lb, c} \approx \theta_{\rm lb, f}$ and
$\cos^2 \theta_{\rm lb, i} \lesssim \p{S / L}_{\rm c0}$,
Eq.~\eqref{eq:angmom_budget_90} can be solved to give
\begin{align}
    \cos \theta_{\rm lb, f} \approx \pm\sqrt{(S / L)_{\rm c0}},
    \label{eq:If_near90}
\end{align}
where the sign is set by the sign of $\cos \theta_{\rm lb, i}$.
Since Eq.~\eqref{eq:If_near90} gives the minimum $\cos\theta_{\rm lb, f}$, we
see that values of $\theta_{\rm lb, f}$ near $90^\circ$ cannot be obtained, an
effect that was identified in \citet{vick2023_HJ} and can be seen in the middle
panel of Fig.~\ref{fig:tau3plots}.

\textbf{(ii)}
When instead $\theta_{\rm lb, i} \ll 1$, $\theta_{\rm sb, f} \leq
90^\circ$, and we can simplify Eq.~\eqref{eq:final_constr1} to obtain:
\begin{align}
    (S / L)_{\rm c} + \cos \theta_{\rm lb, i}
        = \cos \theta_{\rm lb, f} &\leq 1
\end{align}
which gives
\begin{align}
    \theta_{\rm lb, i} &\gtrsim \sqrt{2\p{S / L}_{\rm c}}.
    \label{eq:Ii_near0}
\end{align}
If $\theta_{\rm lb, i}$ is below this bound, then the disk does not have
sufficient angular momentum to tilt the star, and $\theta_{\rm lb, f}$
will be significantly lower than $\theta_{\rm lb, i}$, even nearing $0$,
as can be seen in the middle panel of Fig.~\ref{fig:tau3plots}; the
resulting changes to the obliquity and $\theta_{\rm sb, f}$
distributions are a consequence of this, as shown by the green vertical
dashed lines in all three panels.

Note that the value of $(S / L)_{\rm c}$ in Eq.~\eqref{eq:Ii_near0} is
$\approx (S/L)_{\rm c0}$ as defined in Eq.~\eqref{eq:def_slc}, since
$\cos \theta_{\rm lb, i} \approx 1$ here. Thus, Eq.~\eqref{eq:def_slc}
gives the critical scale of the corrections to the simple $S / L = 0$
result (Section~\ref{ss:th_sllL}) in all discrepant regions.

\textbf{(iii)}
Finally, if $\theta_{\rm lb, i}$ in neither of these two regimes, $\theta_{\rm
lb, f} \approx \theta_{\rm lb, i}$ and the solution is close to the $S = 0$
solution.

\subsection{Final Obliquity Distribution}\label{ss:obl_dist}

Using the above results, we can infer the distribution of present-day
misalignment angles. We must make an assumption about the
initial distribution of disk-binary misalignment angles ($\theta_{\rm lb, i}$),
and we will initially assume an isotropic distribution (uniform in $\cos
\theta_{\rm lb, i}$). Since $\theta_{\rm sb}$ and $\theta_{\rm lb}$ are
approximately constant at late times, their distributions can be
straightforwardly obtained via numerical quadrature of the semi-analytical
results of the previous section, both in the $S \ll L$ limit
(Section~\ref{ss:sol_sllL}) and in the $S \sim L$ case
(Section~\ref{ss:sol_ssimL}).
% i.e.:
% \begin{equation}
%     f(\theta_{\rm sb, f})
%         = \int\limits_{-1}^1
%             f(\theta_{\rm sb, f} | \cos \theta_{\rm lb, i})
%                 \;\mathrm{d} \cos \theta_{\rm lb, i},
% \end{equation}
% and similarly for $f(\theta_{\rm lb, f})$.

\begin{figure}
    \centering
    \includegraphics[width=\columnwidth]{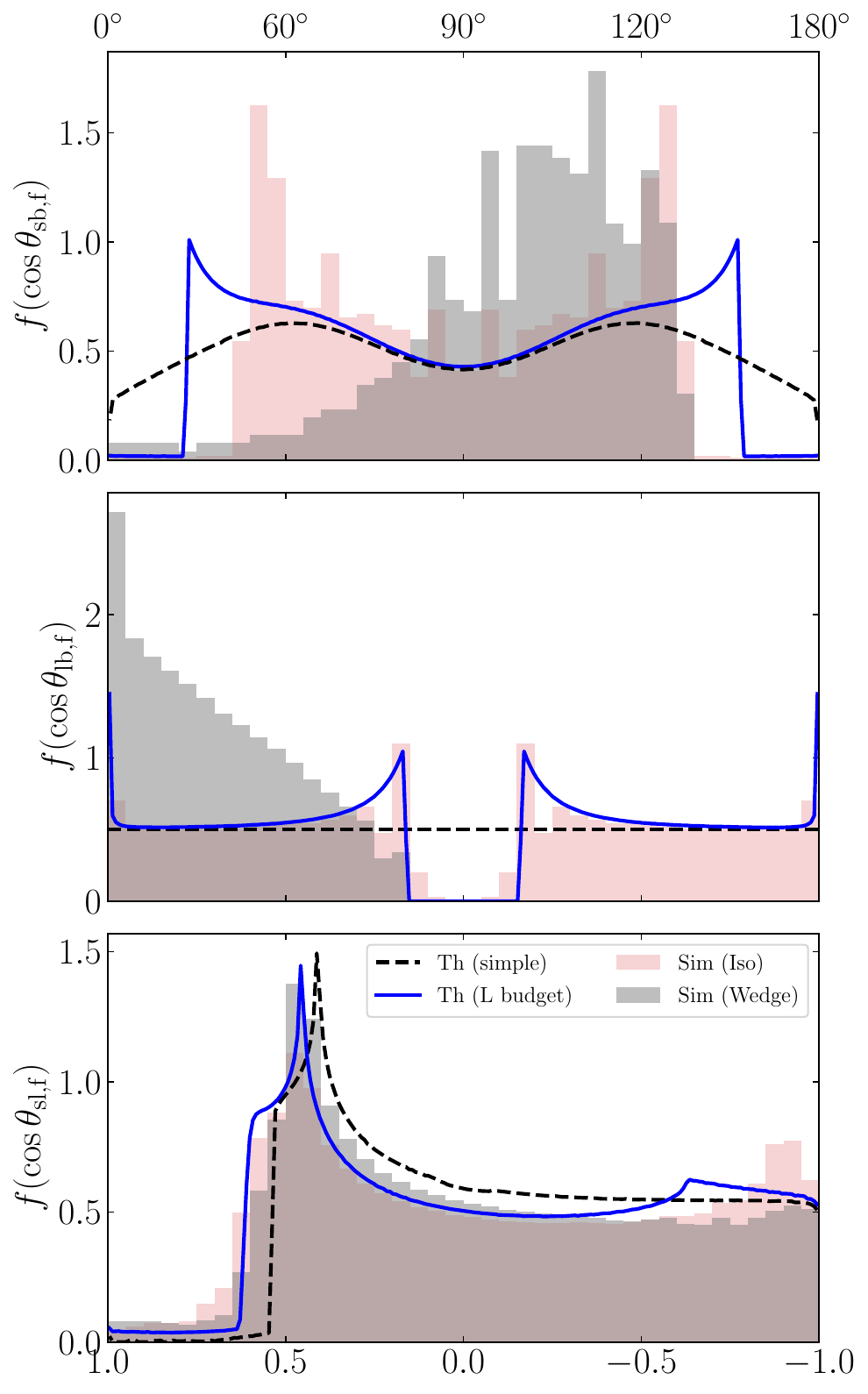}
    \caption{
    The distributions $f$ of the final $\cos \theta_{\rm sb}$, $\cos \theta_{\rm
    lb}$, and $\cos \theta_{\rm sl}$ (the obliquity).
    The red histograms are taken from the numerical integrations shown in
    Fig.~\ref{fig:tau3plots}, the black dashed lines illustrate the theoretical
    prediction in the limit $S \ll L$ (Section~\ref{ss:sol_sllL}), and the blue
    solid lines illustrate the theoretical predictions with the finite $S / L$
    corrections (Section~\ref{ss:sol_ssimL}).
    While all three of these results are given under the assumption of initially
    isotropic disk-binary orientation (uniform distribution of $\cos \theta_{\rm
    lb, i}$), the grey histograms illustrate the numerical distributions when
    assuming a mildly prograde distribution of $\theta_{\rm lb, i}$
    (Eq.~\ref{eq:f_cosqlb_aligned}).
    }\label{fig:histall}
\end{figure}

However, after the disk has fully dissipated, the stellar spin precesses
approximately uniformly about the $\uv{l}_{\rm b}$ axis, implying that the
obliquity is not constant. Instead, $\cos\theta_{\rm sl}$ varies sinusoidally in
time over its geometrically-permitted range (Eq.~\ref{eq:qslf_triangle}). It is
straightforward to show that the probability density function of $\cos
\theta_{\rm sl}$ observed at a random time is
\begin{equation}
    f(x \equiv \cos \theta_{\rm sl})
        \propto \frac{1}{\sqrt{
        (x_{\max} - x)
        (x - x_{\min})
        }}.
\end{equation}
Using this, we can also compute via numerical quadrature the distribution
$f(\theta_{\rm sl, f})$. The distributions of the three angles, and comparisons
to the theoretical results, can be found in Fig.~\ref{fig:histall}.

However, there is evidence for preferential alignment of planetary systems and
their binary companions for $a_{\rm b} \lesssim 500\;\mathrm{au}$, suggesting
that $\theta_{\rm lb, i}$ may not be isotropically distributed \citep[first
pointed out by][]{dupuy2022orbital, christian2022align}.
To account for this possible preference, we also explore the distributions of
these angles by assuming that $\theta_{\rm lb, i}$ is drawn from the
distribution
\begin{equation}
    f(\cos \theta_{\rm lb, i}) =
        \begin{cases}
            2\cos \theta_{\rm lb, i} & \theta_{\rm lb, i} < 90^\circ,\\
            0 & \theta_{\rm lb, i} > 90^\circ.
        \end{cases}
        \label{eq:f_cosqlb_aligned}
\end{equation}
which is a mildly prograde distribution; these are shown in the grey histograms
of Fig.~\ref{fig:histall}.
It can be seen that, while the distributions of the other angles are affected,
the final obliquity distribution (bottom panel) is not significantly changed.

\subsection{Warm Planets and the Laplace Plane Transition}\label{ss:double}

Here, we briefly comment on the distinct dynamics obtained for closer-in planets.
Recall that at sufficiently early times, both $\eta < 1$ and $S \ll L$ due to
the massive disk.
As the disk dissipates, $\eta$ increases to be $\gg 1$, but $S \lesssim L$ is
still satisfied (e.g.\ Fig.~\ref{fig:sample}).
However, as the disk continues to dissipate, the angular momentum of the star
can begin to dominate that of the planet-disk system for sufficiently close-in
planets, i.e.\ $S \gg L$.
Direct examination of Eq.~\eqref{eq:gCS2_cond} shows that a new parameter regime
occurs when $S / L > \eta$ and both quantities are $\gg 1$, under which case the
equilibrium condition reduces to
\begin{align}
    \frac{\sin 2\theta_{\rm sl}}{\sin 2\theta_{\rm lb}} \approx{}&
        \frac{\omega_{\rm lb}}{\omega_{\rm sl}}
            \frac{L}{S} = \frac{\eta}{\cos \theta_{\rm lb}}\frac{L}{S}\nonumber\\
% >>> 1 / (2 * 0.02) * (1 au)^5 / ((300 au)^3 * (2 Rsun)^2) * 100
% 1.070989
        ={}&
            \p{\frac{0.02}{k_{\rm q\star}}}
            \p{\frac{M_{\rm b}}{M_\star}}
            \p{\frac{a_{\rm p}}{1\;\mathrm{au}}}^5
            \p{\frac{a_{\rm b}}{300\;\mathrm{au}}}^{-3}\nonumber\\
        &\times \p{\frac{R_\star}{2R_{\odot}}}^{-2}
            \p{\frac{P_\star}{3\;\mathrm{day}}}^{-2}
            ,\label{eq:laplace_condition}
\end{align}
where in the second line we've evaluated in the late-time limit $M_{\rm d} \to
0$.
If the right-hand side of Eq.~\eqref{eq:laplace_condition} becomes $\ll 1$,
the equilibria of the system must satisfy $\theta_{\rm sl} \approx 0$, $\pi/2$, and
$\pi$ (cf.\ Fig.~\ref{fig:css}).
These solutions are just the familiar Laplace equilibria
\citep{tremaine2009laplace, saillenfest2021future}, the equilibrium orientations
of a test particle (the planet-disk system) subject to two secular precessional
torques (the stellar quadrupole and the binary orbit).
While the detailed evolution in this regime is beyond the scope of this work, we
show here that misalignment can readily be attained and provide a simple
physical explanation.

\begin{figure}
    \centering
    \includegraphics[width=\columnwidth]{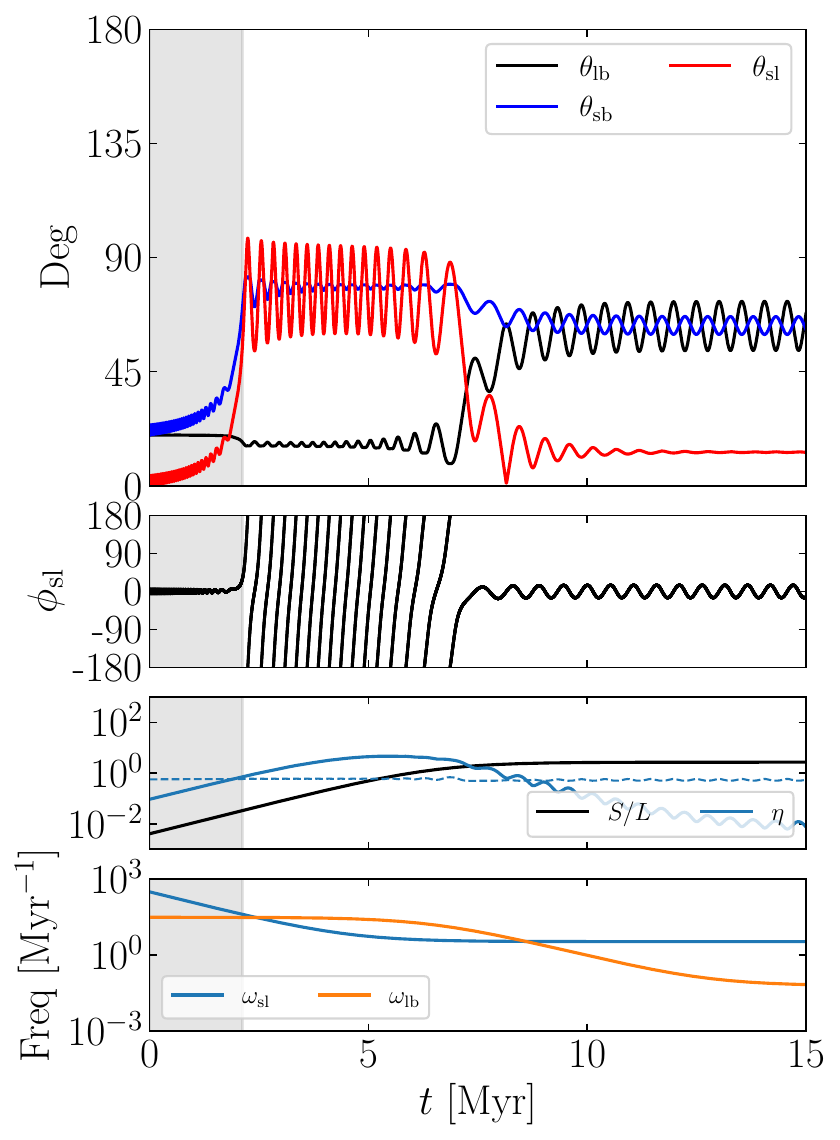}
    \caption{
    Same as Fig.~\ref{fig:sample} except with $a_{\rm p} = 0.5\;\mathrm{au}$.
    The Laplace plane transition occurs when $S / L > \eta \gg 1$ is satisfied,
    leading to a realignment of the planet's orbit with the stellar spin axis.
    }\label{fig:sample_WJ}
\end{figure}

We first illustrate an example of the evolution when a transition to the Laplace
plane regime occurs in Fig.~\ref{fig:sample_WJ}, where the same parameters are
used as those in Fig.~\ref{fig:sample} except that $a_{\rm p} =
0.5\;\mathrm{au}$ is taken.
It can be seen that a realignment of the planet's orbit with the stellar spin
axis (a decrease in $\theta_{\rm sl}$) occurs when
$S / L > \eta$, or when Eq.~\eqref{eq:laplace_condition} decreases below $1$.

\begin{figure}
    \centering
    \includegraphics[width=\columnwidth]{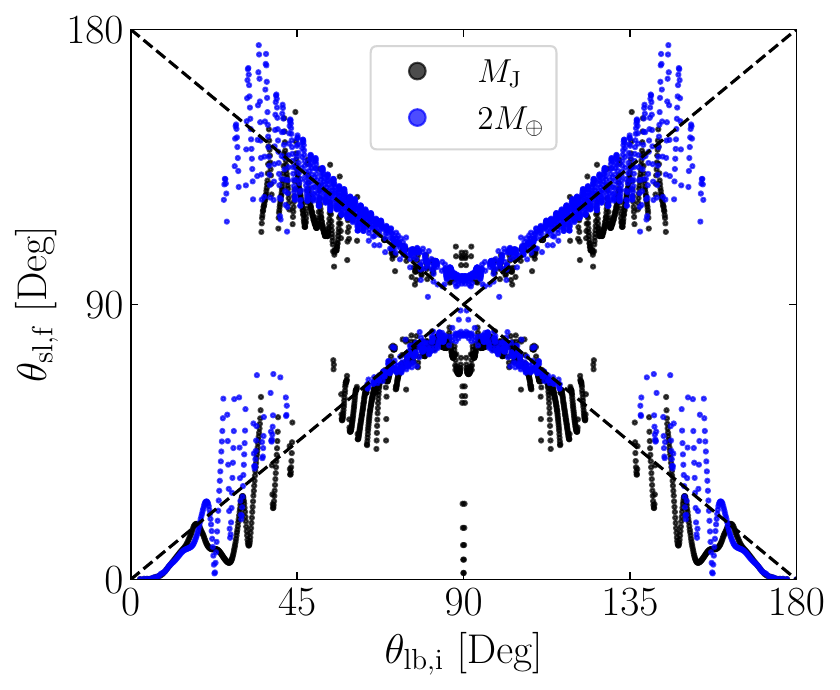}
    \caption{Plots for the final obliquities $\theta_{\rm sl, f}$ as a function of
    the initial $\theta_{\rm lb, i}$ for a Jupiter-mass planet (black) and
    $2M_{\oplus}$-mass planet (blue) located at $a_{\rm p} =
    0.5\;\mathrm{au}$.
    Following the results of Section~\ref{ss:double}, it can be seen that
    $\theta_{\rm sl, f} = \theta_{\rm lb, i}$ and $180^\circ - \theta_{\rm lb,
    i}$ (the two black dashed lines) describe the trend of final obliquities
    well, aside from oscillations that are due to non-adiabatic effects during
    the Laplace plane transition.
    }\label{fig:thetasl_warm}
\end{figure}

The evolution across this transition can be understood simply in a qualitative
manner.
After the stellar spin has been ejected from CS1, $\eta \gg 1$, and the system's
evolution in an inertial frame can be described as rapid precession of $\uv{l}$
about $\uv{l}_{\rm b}$ (with angle $\theta_{\rm lb} \approx \theta_{\rm lb, i}$)
and slow precession of $\uv{s}$ about $\uv{l}_{\rm b}$.
But then, as $L$ continues to decrease, the precession of the disk driven by
the stellar quadrupole will become comparable to, and subsequently exceed, that
driven by the binary companion.
During this transition, the precession axis of $\uv{l}$ must also adiabatically
transition from $\uv{l}_{\rm b}$ to $\pm \uv{s}$, with the sign set by whichever
is closer.
To be explicit, if $\theta_{\rm sb} < 90^\circ$ when the right-hand side of
Eq.~\eqref{eq:laplace_condition} crosses unity, the final precession axis is
$\approx +\uv{s}$, and the final obliquity must be $\theta_{\rm sl, f} =
\theta_{\rm lb, i}$.
On the other hand, if $\theta_{\rm sb} > 90^\circ$, the final obliquity must be
$\theta_{\rm sl, f} = 180^\circ - \theta_{\rm lb, i}$.
Figure~\ref{fig:thetasl_warm} shows the final obliquities for a range of
$\theta_{\rm lb, i}$ for both a Jupiter-mass planet (black) and super Earth-mass
planet (blue) located at $a_{\rm p} = 0.5\;\mathrm{au}$.
As described by Eq.~\eqref{eq:laplace_condition}, the system dynamics transition
to those of the Laplace plane for such values of $a_{\rm p}$, and the final
outcome is roughly independent of planet mass.

\subsection{Physical Parameter Space of Obliquity
Excitation}\label{ss:param_space}

\begin{figure}
    \centering
    \includegraphics[width=\columnwidth]{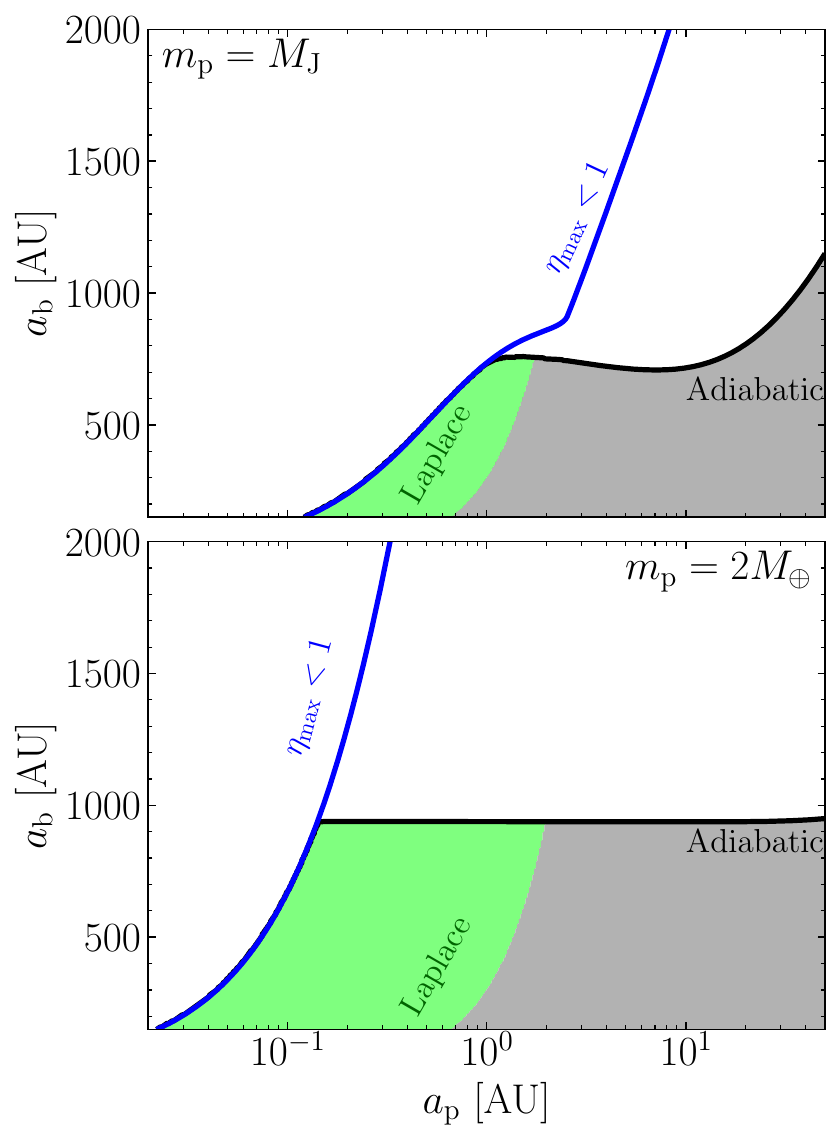}
    \caption{
    Parameter space that are relevant for the resonant obliquity excitation,
    which requires that the ratio $\eta$ (Eq.~\ref{eq:def_eta}) cross
    $1$ adiabatically, for a Jupiter-mass planet (top) and a $2M_\oplus$ planet
    (bottom).
    The blue lines denote $\eta_{\max} = 1$, and the black line denotes the
    adiabaticity constraint (Eq.~\ref{eq:constr_adiabatic}); note that
    $\eta_{\rm i} < 1$ for the entire parameter space shown.
    The grey shaded regions are where obliquity excitation following the results
    of Section~\ref{ss:sol_ssimL} are expected.
    The green shaded region in the top panel corresponds to a regime where
    resonance crossing occurs even though $\eta_{\rm i} < 1$ and $\eta_{\rm f} <
    1$ (see Section~\ref{ss:double}).
    The parameters held fixed are: $M_{\rm d, i} = 0.1M_\odot$, $M_\star =
    M_{\rm b} = M_{\odot}$, $r_{\rm in} = 4R_{\star} = 8R_\odot$, $r_{\rm out} =
    50\;\mathrm{au}$, and $\Omega_\star / \sqrt{GM_\star/R_\star^3} = 0.1$.
    The lowest value of $a_{\rm b}$ shown is $150\;\mathrm{au} = 3r_{\rm out}$,
    below which truncation of the protoplanetary disk is expected
    \citep{artymowicz1994disk, miranda2015disk}.
}\label{fig:param_space}
\end{figure}

Finally, we discuss the regions of physical parameter space where our analytical
and semi-analytical results above may be expected to be accurate.
Despite the numerous free parameters of the problem describing the properties of
the planet, disk, and binary, the essential features of the problem simply
require that the system evolve adiabatically from $\eta_{\rm i} > \eta_{\rm c}
\sim 1$ to $\eta_{\rm f} < \eta_{\rm c} \sim 1$
(Eqs.~\ref{eq:def_etac0},~\ref{eq:eta_i},~\ref{eq:eta_f}).
The adiabaticity constraint simply requires
\begin{equation}
    \s{\omega_{\rm lb}}_{\eta \simeq 1} \tau_{\rm d} \gg 1.
        \label{eq:constr_adiabatic}
\end{equation}
In Fig.~\ref{fig:param_space}, we show in grey the regions of
phase space where the canonical results derived in Section~\ref{ss:obl_dist} are
expected due to sufficiently adiabatic evolution.
We also indicate in green the regions where the Laplace plane transition occurs
(Section~\ref{ss:double}), where obliquity excitation is also expected, but
following a qualitatively different distribution and mechanism.
In agreement with
\citet{zanazzi2018Paper1}, we find that no
obliquity is excited for a Jupiter-mass planet with $a_{\rm p}
\lesssim 0.25\;\mathrm{au}$ if $a_{\rm b} = 300\;\mathrm{au}$, because $\eta$
remains below $1$ throughout the system evolution.
For the same reason, no obliquity is excited for an Earth-mass planet with
$a_{\rm p} \lesssim 0.03\;\mathrm{au}$.

\section{Obliquity Evolution with A Broken Disk}\label{s:broken_disk}

The discussion given above in Section~\ref{s:unbroken_dynamics} and summarized
in Fig.~\ref{fig:param_space} suggests that the mass of a planet has a
significant effect on the region of parameter space where obliquity excitation
is expected.
In particular, super-Earths with $a_{\rm p} \simeq 0.1\;\mathrm{au}$ can readily
become misaligned from their host stars, in tension with the observed good
alignment of super-Earth systems \citep{albrecht2022review}.
% However, observational evidence suggests that the misalignment of planetary
% orbits is not especially sensitive either to the planet's mass or orbital
% separation \citep{albrecht2022review, louden2024}.
While our model predictions cannot be compared at face value to the
observational results, which are not restricted to planets with binary
companions, this inconsistency motivates our next consideration,
i.e.\ non-homologous disk dissipation.

We consider here the standard picture where a protoplanetary disk (PPD)
experiencing photoevaporative winds evolves through a transition disk phase.
In this picture, after the mass accretion rate falls to below the the
photoevaporative mass loss rate, a gap at $\sim 2\;\mathrm{au}$ develops.
As this gap cuts off the supply of mass to the inner regions of the PPD, the PPD
interior to $\sim 2\;\mathrm{au}$ clears out on a timescale $\sim
10^5\;\mathrm{yr}$, and then the remainder of the disk slowly disappears on the
standard PPD lifetime $\sim 10^6\;\mathrm{yr}$ (e.g.\ see
\citealp{alexander2014review} for a theoretical review and
\citealp{vandermarel2016_whole} for a more recent observational sample of
transition disks largely consistent with the values we adopt).
While more recent works suggest that magnetohydrodynamic winds may affect the
formation of this cavity \citep[e.g.][]{suzuki2016mhdwind, kunitomo2020mhd}, the
dynamical consequences of a PPD gap opened by photoevaporation significantly
complicates the scenario described in
Sections~\ref{s:unbroken_theory}--\ref{s:unbroken_dynamics}.
In these sections, we assumed that the disk evolves as a nearly rigid plane,
since warps are rapidly communicated at the sound-crossing time via bending
waves \citep[e.g.][]{foucart2014_warp, zanazzi2018Paper2, gerbig2024}.
However bending waves cannot propagate through a gap.
Thus, if the PPD contains a gap, then it must instead be treated as two
hydrodynamically independent, though gravitationally coupled, disks.

In this section, we will explore some of the interesting outcomes of this model,
where there are two stars, two disks, and a planet.
Owing to the numerous free parameters and uncertainties, a complete
characterization of these dynamics is beyond the scope of this paper, but we
demonstrate a few interesting effects exclusive to this model.

\subsection{Disk Model and Equations of Motion}\label{ss:twodisk_eom}

We will consider a system with similar parameters to those explored in
Section~\ref{s:unbroken_dynamics}, but instead with an inner disk extending from
$r_{\rm i, in} = 8R_\odot$ to $r_{\rm i, out} = 1\;\mathrm{au}$, and an outer
disk extending from $r_{\rm o, in} = 3\;\mathrm{au}$ to $r_{\rm o, out} =
50\;\mathrm{au}$.
Here, the notation $r_{\rm i, out}$ denotes the inner (``i'') disk's outer
(``out'') radius, and correspondingly for each of $r_{\rm i, in}$, $r_{\rm o,
in}$, and $r_{\rm o, out}$.
We will further consider that the planet, which we take to be a super Earth
initially, is embedded in the inner disk---we argue later that planets embedded
in the outer disk will result in obliquity distributions similar to those from
the previous sections.
The evolution of such a system can be described by
\begin{align}
    \rd{\uv{s}}{t} &=
        - \omega_{\rm sl}
            \p{\uv{l} \cdot \uv{s}}\p{\uv{l} \times \uv{s}},\label{eq:dsdt_2d}\\
    \omega_{\rm sl} &= \omega_{\rm sp} + \omega_{\rm si},\nonumber\\
    \rd{\uv{l}}{t}
        &= \omega_{\rm ls}
            \p{\uv{l} \cdot \uv{s}}\p{\uv{l} \times \uv{s}}
        - \omega_{\rm lo}
                \p{\uv{l}_{\rm o} \cdot \uv{l}}
                \p{\uv{l}_{\rm o} \times \uv{l}},\\
    \omega_{\rm ls} &\equiv \omega_{\rm sl}\frac{S}{L},\nonumber\\
    \omega_{\rm lo} &= \omega_{\rm po}\frac{L_{\rm p}}{L}
                + \omega_{\rm io}\frac{L_{\rm i}}{L},\nonumber\\
    \rd{\uv{l}_{\rm o}}{t}
        &= \omega_{\rm lo}\frac{L}{L_{\rm o}}
                \p{\uv{l}_{\rm o} \cdot \uv{l}}
                \p{\uv{l}_{\rm o} \times \uv{l}}
        - \omega_{\rm ob} \p{\uv{l}_{\rm b} \cdot
                \uv{l}_{\rm o}}\p{\uv{l}_{\rm b} \times \uv{l}_{\rm o}}.
\end{align}
Here, $L_{\rm p}$ and $L_{\rm i}$ are the magnitudes of the angular momenta of
the planet and inner disk, $\bm{L} = \bm{L}_{\rm p} + \bm{L}_{\rm i} = L\uv{l}$ is the
orientation of their combined angular momentum, and $L_{\rm o}$ and $\uv{l}_{\rm
o}$ are the angular momentum magnitude and orientation of the outer disk.
We neglect the torque from the binary companion on the inner disk, which is
justified since $r_{\rm i, out} \ll a_{\rm b}$.
We take $\uv{s} \parallel \uv{l} \parallel \uv{l}_{\rm o}$ initially.
The precession frequencies are:
\begin{align}
    \omega_{\rm si} ={}& \frac{3k_{\rm q\star}}{4k_\star}
            \p{\frac{M_{\rm i}}{M_\star}}
            \p{\frac{R_\star^3}{r_{\rm i, in}^2r_{\rm i, out}}}
            \Omega_\star\nonumber\\
        ={}& \frac{2\pi}{38\;\mathrm{kyr}}
            \p{\frac{2k_{\rm q\star}}{k_\star}}
            \p{\frac{M_{\rm i}}{10^{-3}M_\star}}
            \p{\frac{r_{\rm i, in}}{8 R_{\odot}}}^{-2}\nonumber\\
        &\times \p{\frac{r_{\rm i, out}}{1\;\mathrm{au}}}^{-1}
            \p{\frac{P_\star}{3 \;\mathrm{days}}}^{-1}
            \p{\frac{R_\star}{2R_{\odot}}}^{3},
% 2  * pi / (yr * 3/8 / 10^(3) * ((2 Rsun)^3 / ((8 Rsun)^2 * au)) * (2 * pi / (3 day)))
% 3.769086e+04
\end{align}
\begin{align}
    \omega_{\rm io} ={}& \frac{3M_{\rm o}}{16M_\star}
            \p{\frac{r_{\rm i, out}^3}{r_{\rm o, in}^2r_{\rm o, out}}}
            n_{\rm i, out} \nonumber\\
        ={}& \frac{2\pi}{24\;\mathrm{kyr}}
            \p{\frac{M_{\rm o}}{0.1 M_\odot}}
            \p{\frac{M_\star}{M_{\odot}}}^{-1/2}
            \p{\frac{r_{\rm i, out}}{1 \;\mathrm{au}}}^{3/2}\nonumber\\
        &\times \p{\frac{r_{\rm o, in}}{3 \;\mathrm{au}}}^{-2}
            \p{\frac{r_{\rm o, out}}{50\;\mathrm{au}}}^{-1},\label{eq:def_wio}
% 2 * pi / (yr * (3/16 * 0.1 * (1 au)^3 / ((3 au)^2 * (50 au)) * (G * Msun / (au)^3)^(1/2)))
% 2.399860e+04
\end{align}
\begin{align}
    \omega_{\rm ob} ={}& \frac{3M_{\rm b}}{8M_\star}
            \p{\frac{r_{\rm o, out}}{a_{\rm b}}}^3 n_{\rm o, out}\nonumber\\
        ={}& \frac{2\pi}{0.2\;\mathrm{Myr}}
            \p{\frac{M_{\rm b}}{M_\star}}
            \p{\frac{M_\star}{M_{\odot}}}^{1/2}
            \p{\frac{r_{\rm o, out}}{50\;\mathrm{au}}}^{3/2}\nonumber\\
            &\times \p{\frac{a_{\rm b}}{300\;\mathrm{au}}}^{-3},
% 2 * pi / (yr * 3/8 * ((50 au) / (300 au))^3 * (G * (Msun) / (50 au)^3)^(1/2))
% 2.036349e+05
\end{align}
\begin{align}
    \omega_{\rm sp} ={}& \frac{3k_{\rm q\star}}{2k_\star}
            \p{\frac{m_{\rm p}}{M_\star}}
            \p{\frac{R_\star}{a_{\rm p}}}^3 \Omega_\star\nonumber\\
        ={}& \frac{2\pi}{18\;\mathrm{Myr}}
            \p{\frac{2k_{\rm q\star}}{k_\star}}
            \p{\frac{m_{\rm p}}{2 M_{\oplus}}}
            \p{\frac{M_\star}{M_{\odot}}}^{-1}\nonumber\\
        &\times \p{\frac{R_\star}{2R_{\odot}}}^{3}
            \p{\frac{a_{\rm p}}{0.2\;\mathrm{au}}}^{-3}
            \p{\frac{P_\star}{3\;\mathrm{days}}}^{-1},
% 2 * pi / (yr * 3 / 4 * (2 Mearth) / Msun * ((2 Rsun) / (0.2 au))^3 * (2 * pi / (3 day)))
% 1.814972e+07
\end{align}
\begin{align}
    \omega_{\rm po} ={}& \frac{3M_{\rm o}}{8M_\star}
            \p{\frac{a_{\rm p}^3}{r_{\rm o, in}^2r_{\rm o, out}}}
            n_{\rm p}\nonumber\\
        ={}& \frac{2\pi}{0.18\;\mathrm{Myr}}
            \p{\frac{M_{\rm o}}{0.1 M_\odot}}
            \p{\frac{M_\star}{M_{\odot}}}^{-1/2}\nonumber\\
        &\times \p{\frac{a_{\rm p}}{0.2 \;\mathrm{au}}}^{3/2}
            \p{\frac{r_{\rm o, in}}{3 \;\mathrm{au}}}^{-2}
            \p{\frac{r_{\rm o, out}}{50\;\mathrm{au}}}^{-1}.
% 2 * pi / (yr * (3/8 * 0.1 * (0.2 au)^3 / ((3 au) * (50 au)^2) * (G * Msun / (0.2 au)^3)^(1/2)))
% 2.235938e+06
\end{align}
These expressions are easily derived/adapted from previous works
\citep{lai2014star} except for Eq.~\eqref{eq:def_wio}, which is derived in
Appendix~\ref{app:wio}.
The angular momentum ratios are
\begin{align}
    \frac{S}{L_{\rm p}} ={}& \frac{k_\star M_\star R_\star^2 \Omega_\star}{
            m_{\rm p} \sqrt{GM_\star a_{\rm p}}}\nonumber\\
        ={}& 392
            \p{\frac{M_\star}{M_{\odot}}}^{1/2}
            \p{\frac{k_\star}{0.1}}
            \p{\frac{R_\star}{2 R_{\odot}}}^2\nonumber\\
        &\times \p{\frac{P_\star}{3\;\mathrm{day}}}^{-1}
            \p{\frac{m_{\rm p}}{2M_{\oplus }}}^{-1}
            \p{\frac{a_{\rm p}}{0.2\;\mathrm{au}}}^{-1/2},
\end{align}
% 0.1 * Msun * (2 Rsun)^2 / (2 * Mearth * (G * Msun * (0.2 au))^(1/2)) * (2 * pi / (3 day))
% 391.915854
\begin{align}
    \frac{S}{L_{\rm i}} ={}& 2\frac{k_\star M_\star R_\star^2 \Omega_\star}{
            M_{\rm i} \sqrt{GM_\star r_{\rm i, out}}}\nonumber\\
        ={}& 2
            \p{\frac{M_\star}{M_{\odot}}}^{1/2}
            \p{\frac{k_\star}{0.1}}
            \p{\frac{R_\star}{2 R_{\odot}}}^2\nonumber\\
        &\times \p{\frac{P_\star}{3\;\mathrm{day}}}^{-1}
            \p{\frac{M_{\rm i}}{10^{-3}M_{\star}}}^{-1}
            \p{\frac{r_{\rm i, out}}{\;\mathrm{au}}}^{-1/2},
% 0.1 * (2 Rsun)^2 / (1e-3 / 2 * (G * Msun * (au))^(1/2)) * (2 * pi / (3 day))
% 2.105004
\end{align}
\begin{align}
    \frac{S}{L_{\rm o}} ={}& 2\frac{k_\star M_\star R_\star^2 \Omega_\star}{
            M_{\rm o} \sqrt{GM_\star r_{\rm out}}}\nonumber\\
        ={}& \scinot{8.93}{-3}
            \p{\frac{M_\star}{M_{\odot}}}^{1/2}
            \p{\frac{k_\star}{0.1}}
            \p{\frac{R_\star}{2 R_{\odot}}}^2\nonumber\\
        &\times \p{\frac{P_\star}{3\;\mathrm{day}}}^{-1}
            \p{\frac{M_{\rm o}}{0.1M_{\star}}}^{-1}
            \p{\frac{r_{\rm out, o}}{50\;\mathrm{au}}}^{-1/2}.
% 0.1 * (2 Rsun)^2 / (0.1 / 2 * (G * Msun * (50 au))^(1/2)) * (2 * pi / (3 day))
% 0.008931
\end{align}
The masses of the two disks are $M_{\rm i}$ and $M_{\rm o}$.
We assume that both disks dissipate homologously, so that
\begin{align}
    M_{\rm i}(t) &= M_{\rm i, i}e^{-t / \tau_{\rm d, i}} &
    M_{\rm o}(t) &= M_{\rm o, i}e^{-t / \tau_{\rm d, o}}.
\label{eq:def_taudido}
\end{align}
As described above, we take $\tau_{\rm d, i} = 0.1\;\mathrm{Myr}$ and $\tau_{\rm
d, o} = 1\;\mathrm{Myr}$ \citep{alexander2014review}.
For the initial disk
masses, we take $M_{\rm i, i} = 10^{-3}M_\star$ (sufficient material to form
Jupiter-mass planets interior to $\sim 1\;\mathrm{au}$) and $M_{\rm o, i} =
0.1M_{\star}$ as fiducial parameters.
Physically, this corresponds to efficient
photoevaporative winds that almost immediately open a gap in a $0.1M_\star$
disk; such rapid gap opening is unlikely to be realistic, and we explore more
realistic gap-opening times in Section~\ref{ss:gap_timing}.

\subsection{Evolution and Obliquity Excitation}\label{ss:twop_analysis}

\begin{figure*}
    \centering
    \includegraphics[width=\columnwidth]{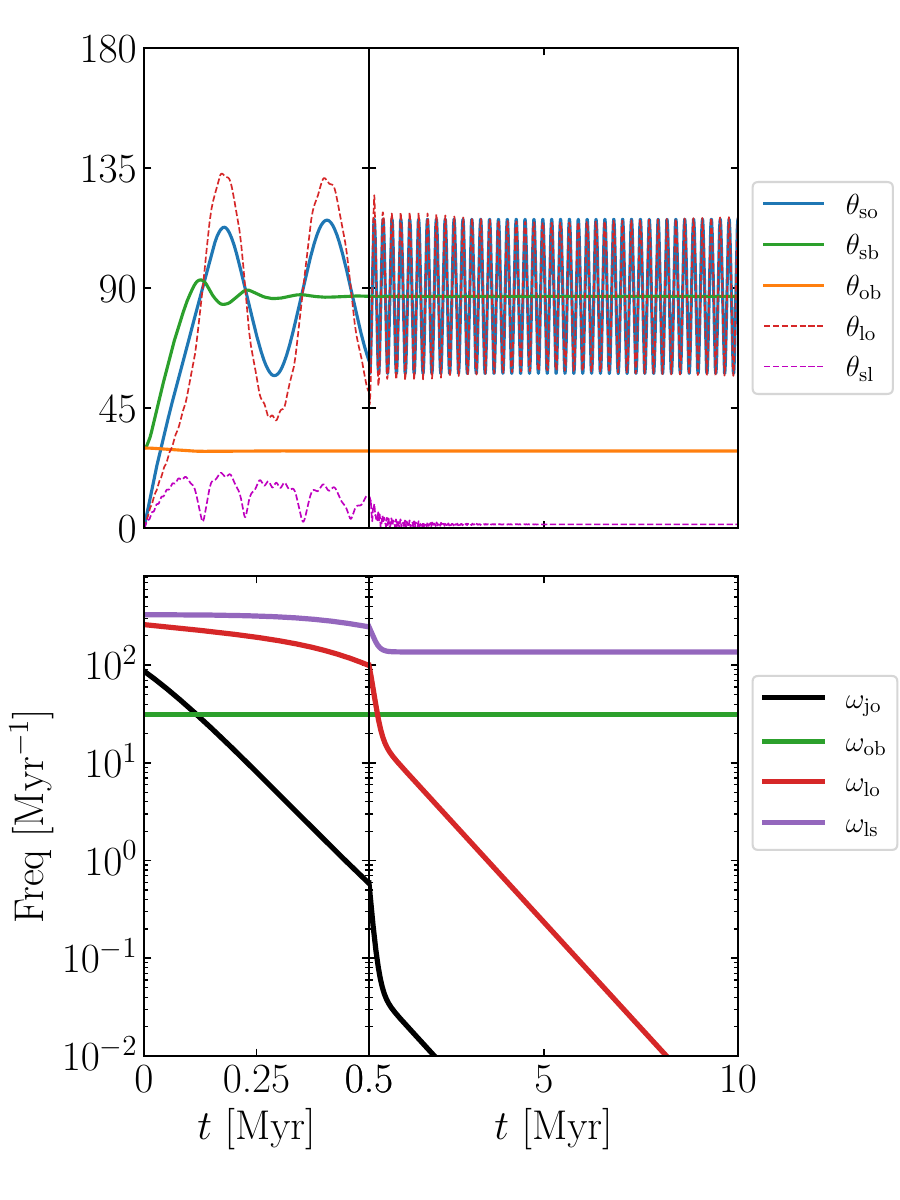}
    \includegraphics[width=\columnwidth]{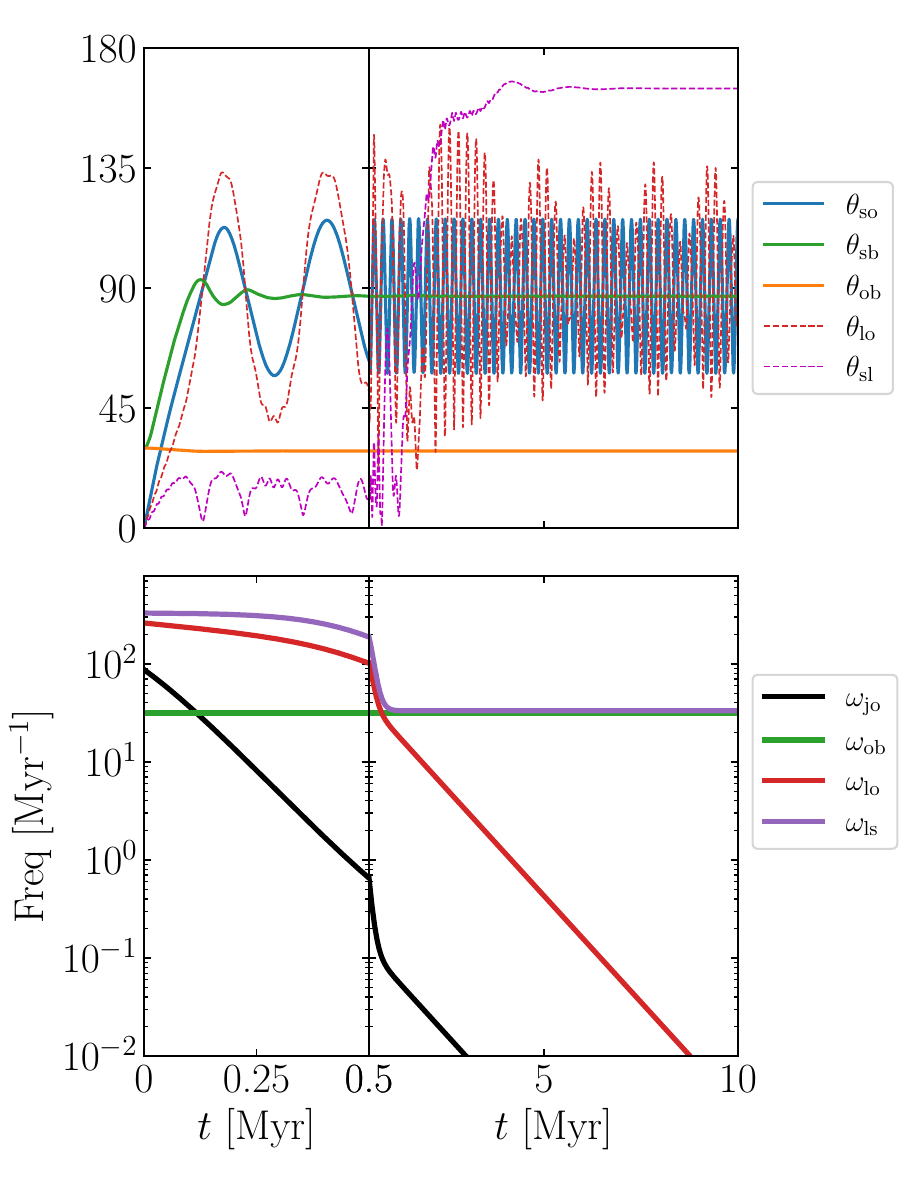}
    \caption{
    Left: The top panel shows the evolution of the misalignment angles
    between the spin (s), planet+inner disk (l), outer disk (o), and binary (b)
    for a $2M_\oplus$ planet located at $0.2\;\mathrm{au}$.
    It can be seen that an early misalignment between the stellar spin and outer
    disk ($\theta_{\rm so}$, blue solid line) is excited, and a small obliquity
    between the spin and planet+inner disk ($\theta_{\rm sl}$; magenta dashed
    line) develops, but once both disks have fully dissipated, the obliquity goes
    to zero.
    By comparing with the bottom panel, we see that the initial tilting of the
    stellar spin is due to the crossing of the $\omega_{\rm jo} \simeq
    \omega_{\rm ob}$ commensurability, where $\omega_{\rm jo}$ is the precession
    frequency of the combined star-inner disk-planet system about the outer disk
    (Eq.~\ref{eq:def_omega_jo}).
    The solid lines are related to the outer disk orientation, and correspond
    closely to the solid lines in Fig.~\ref{fig:sample}, while the dashed lines
    are related to the inner disk orientation, and are new to the broken disk
    mechanism.
    Right: The same but for a planet semimajor axis of $0.3\;\mathrm{au}$, where a
    large obliquity $\theta_{\rm sl}$ is attained after both disks have
    dissipated.
    This is because $\omega_{\rm ls} \gg \omega_{\rm lo}$
    (Eq.~\ref{eq:2d_res_jump}) is not sufficiently well satisfied at $t \simeq
    1\;\mathrm{Myr}$, introducing misalignment at this time, as can be seen
    in the top panel.
    In both cases, misalignment of the inner and outer disks (finite
    $\theta_{\rm lo}$) is essential to the dynamics.
    }\label{fig:sample_SE}
\end{figure*}

Figure~\ref{fig:sample_SE} shows the integration of
Eqs.~(\ref{eq:dsdt_2d}--\ref{eq:def_taudido}) for two different values of the
planetary semimajor axis.
There are two important transitions during the system's evolution:
\begin{itemize}
    \item Early on, the coupling between the stellar spin and inner disk is the
        strongest in the system, as the driving by the outer disk is
        insufficient to break alignment: following \citet{lai2018misalign}, we
        use the fiducial parameters to evaluate the spin-inner disk coupling
        strength
        \begin{equation}
            \epsilon_{\rm sl}
                = \frac{\omega_{\rm lo}}
                    {\omega_{\rm sl} + \omega_{\rm ls}}
                \approx 0.5.\label{eq:eq50}
        \end{equation}
        Note that if $\epsilon_{\rm sl} \ll 1$, then the spin and inner disk are
        strongly coupled (we have evaluated Eq.~\ref{eq:eq50} assuming that the
        inner disk+planet are dominated by the disk, i.e.\ $\omega_{\rm sl}
        \approx \omega_{\rm si}$).
        Even though $\epsilon_{\rm sl}$ is not very small, we nevertheless
        approximate that the inner disk and stellar spin can be treated as a
        combined angular momentum $\bm{J} = J\uv{\textbf{\j}}$ obeying:
        \begin{align}
            \rd{\uv{\textbf{\j}}}{t} ={}& -\omega_{\rm jo}
                \p{\uv{l}_{\rm o} \cdot \uv{\textbf{\j}}}
                    \p{\uv{l}_{\rm o} \times \uv{\textbf{\j}}},\\
                \omega_{\rm jo} &\equiv
                    \frac{\omega_{\rm io}L_{\rm i} + \omega_{\rm po} L_{\rm
                    p}}{S + L}\label{eq:def_omega_jo}.
        \end{align}

        The evolution of the three vectors $\uv{\textbf{\j}}$, $\uv{l}_{\rm o}$,
        and $\uv{l}_{\rm b}$ then can be described using the theory laid out
        above in Sections~\ref{s:unbroken_theory}--\ref{s:unbroken_dynamics}.
        Thus, as the inner disk dissipates, $\uv{\textbf{\j}}$ will become
        misaligned with respect to the outer disk as $\omega_{\rm jo}$ crosses
        $\omega_{\rm ob}$ due to dissipation of the inner disk.

        This is why planets embedded in the outer disk (with semimajor axes
        $\gtrsim r_{\rm o, in}$) are expected to present similar obliquities
        whether a photoevaporatively opened gap appears or not: the misalignment
        of an outer disk with respect to the stellar spin evolves similarly as
        does the misalignment of a single, rigid PPD to the stellar spin. This
        is most evident when comparing the evolution of the spin-outer disk
        misalignment angle $\theta_{\rm so}$ in the upper panel of
        Fig.~\ref{fig:sample_SE} to the evolution of the spin-disk misalignment
        $\theta_{\rm sl}$ in the single-disk scenario (Fig.~\ref{fig:sample}).

    \item Later, as the inner disk dissipates, the inner disk+planet system
        transitions to becoming planet-dominated. During this evolution, a
        second resonance can be crossed. Since the star's angular momentum now
        dwarfs that of the inner system, the only vectors that evolve during
        this phase are:
        \begin{align}
            \rd{\uv{l}}{t}
                &= \omega_{\rm ls}
                    \p{\uv{l} \cdot \uv{s}}\p{\uv{l} \times \uv{s}}
                - \omega_{\rm lo}
                        \p{\uv{l}_{\rm o} \cdot \uv{l}}
                        \p{\uv{l}_{\rm o} \times
                        \uv{l}},\label{eq:dldt_resonance}\\
            \rd{\uv{l}_{\rm o}}{t}
                &\approx
                - \omega_{\rm ob} \p{\uv{l}_{\rm b} \cdot
                        \uv{l}_{\rm o}}\p{\uv{l}_{\rm b} \times \uv{l}_{\rm o}}.
        \end{align}
        The precession of $\uv{l}_{\rm o}$ introduces a periodic forcing (with
        frequency $2\omega_{\rm ob}$,
        because $\uv{l}_{\rm o}$ varies with frequency $\omega_{\rm ob}$
        and enters quadratically into the second term of
        Eq.~\ref{eq:dldt_resonance}) into the precession of $\uv{l}$.
        This forcing can resonantly increase $\theta_{\rm sl}$
        if its frequency is resonant with the natural precession
        frequency $\omega_{\rm ls}$ of $\uv{l}$ about $\uv{s}$
        \begin{equation}
            2\omega_{\rm ob} \simeq \omega_{\rm ls} \approx \omega_{\rm ps},
            \label{eq:2d_res_cross}
        \end{equation}
        and the magnitude of the excitation is set by the forcing strength
        \begin{align}
            \Delta \theta_{\rm sl}
                \propto
                \p{\frac{\omega_{\rm lo}}{\omega_{\rm ls}}}_{\rm cross}
                \approx
                \p{\frac{\omega_{\rm po}}{\omega_{\rm ps}}}_{\rm cross}
                    \label{eq:2d_res_jump},
        \end{align}
        where the forcing strength is dominated by the strength at resonance
        crossing, since $\omega_{\rm po} \propto M_{\rm o}$ falls off as the
        outer disk continues to photoevaporate.
\end{itemize}
Note that, since
\begin{align}
    \omega_{\rm ps}
        &= \frac{3k_{\rm q\star}}{2}\p{\frac{R_\star}{a_{\rm p}}}^5
            \frac{\Omega_\star^2}{n_{\rm p}},
\end{align}
does not depend on the planet mass, both the resonance crossing
condition Eq.~\eqref{eq:2d_res_cross} and the amplitude of the
excitation Eq.~\eqref{eq:2d_res_jump} are expected to be independent
of the planet mass, though not its  semimajor axis. As such, our
analysis predicts that the obliquity distribution of planets embedded in
the inner disk depends on the location of the planet, but not its mass.

\begin{figure*}
    \centering
    \includegraphics[width=0.77\columnwidth]{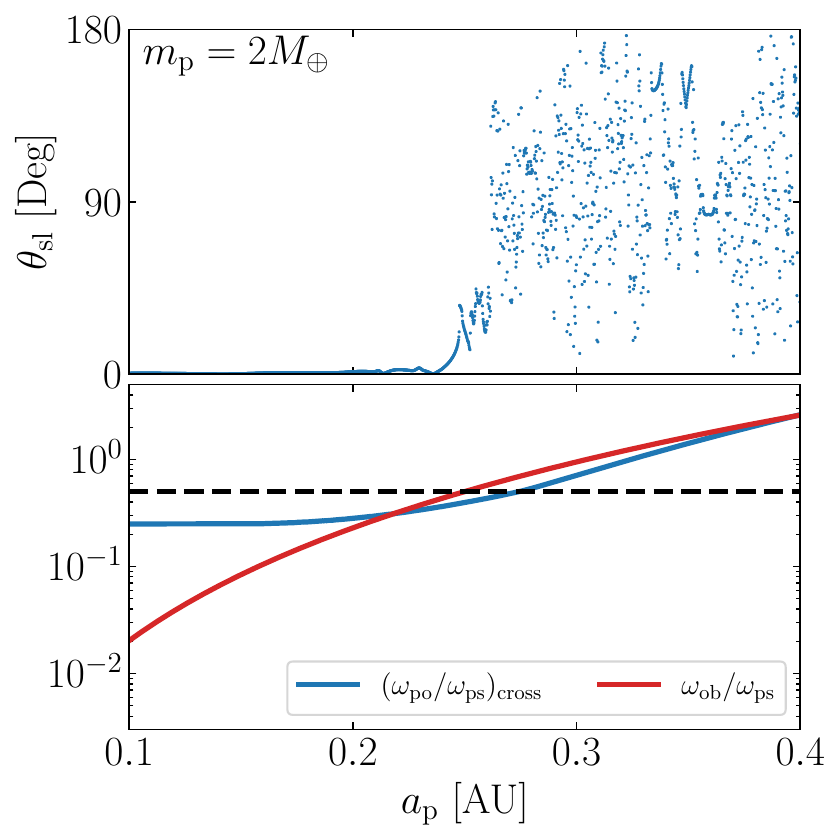}
    \includegraphics[width=0.76\columnwidth]{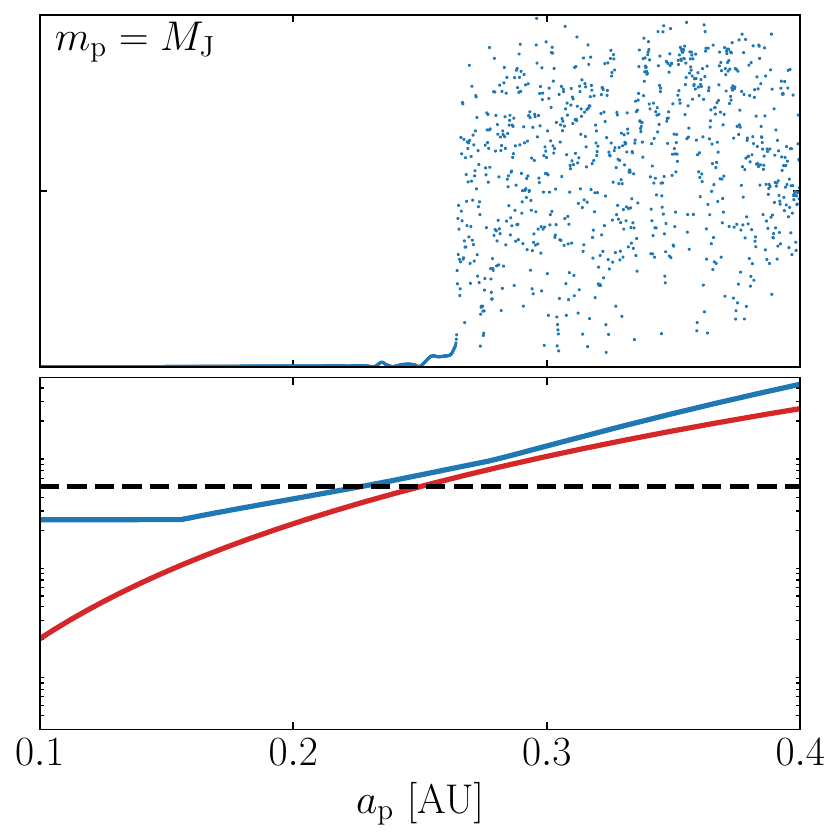}
    \caption{
    Left: The top panel shows the final obliquity $\theta_{\rm sl}$ for a
    $2M_\oplus$-mass planet as a function of the planet's semimajor axis when
    assuming the two-disk evolution model described in
    Section~\ref{ss:twodisk_eom}.
    The bottom panel shows several critical precession frequency ratios as
    obtained from the analysis in Section~\ref{ss:twop_analysis}: it can be seen
    that large obliquities are obtained when $\omega_{\rm ob} \gtrsim
    \omega_{\rm ps} / 2$ (black dashed line; Eq.~\ref{eq:2d_res_cross}), and the
    resonant forcing strength ($\propto \omega_{\rm lo} / \omega_{\rm ls}$ at
    resonance crossing; Eq.~\ref{eq:2d_res_jump}) is not too small.
    Right: Same but for a $M_{\rm J}$-mass planet. The transition between
    alignment and misalignment is at a very similar semimajor axis to the
    $2M_\oplus$ case, supporting the conclusion of
    Section~\ref{ss:twop_analysis} that misalignment in the two-disk scenario is
    not very sensitive to planet mass for planets embedded in the inner disk.
    Initial disk masses of $M_{\rm i, i} = 10^{-3}M_\star$ and $M_{\rm o, i} =
    0.1M_\star$ are used.
    }\label{fig:scanSEWJ}
\end{figure*}

To confirm this prediction, we display the final obliquity as a function of
$a_{\rm p}$ for a $2M_\oplus$-planet and a $M_{\rm J}$-planet in the two columns
of Fig.~\ref{fig:scanSEWJ}, where all other parameters have been set to their
fiducial values.
It can be seen that obliquity excitation indeed occurs when $\omega_{\rm ob}
\gtrsim \omega_{\rm ps} / 2$, and the result does not depend sensitively on the
planet mass.

\subsection{Timing of Gap Opening}\label{ss:gap_timing}

Our fiducial parameters above implicitly assume that the initially
$0.1M_\star$ PPD immediately exhibits a gap due to photoevaporation.
This is unlikely to be realistic: this gap can only be opened once the
photoevaporative mass loss rate exceeds the accretion rate onto the central
star, which typically takes $\sim$ a few Myr \citep[e.g.][]{alexander2014review,
kunitomo2021ppd}.
As for the effect that the timing of this gap opening has on the obliquity
evolution of the central star, the key difference depends on whether the secular
precession resonance takes place before or after the PPD opens a gap.
If resonance crossing has already occurred by the time that a gap has opened,
the results of Section~\ref{s:unbroken_dynamics} can be used to understand the
conditions for obliquity excitation.

\begin{figure*}
    \centering
    \includegraphics[width=0.77\columnwidth]{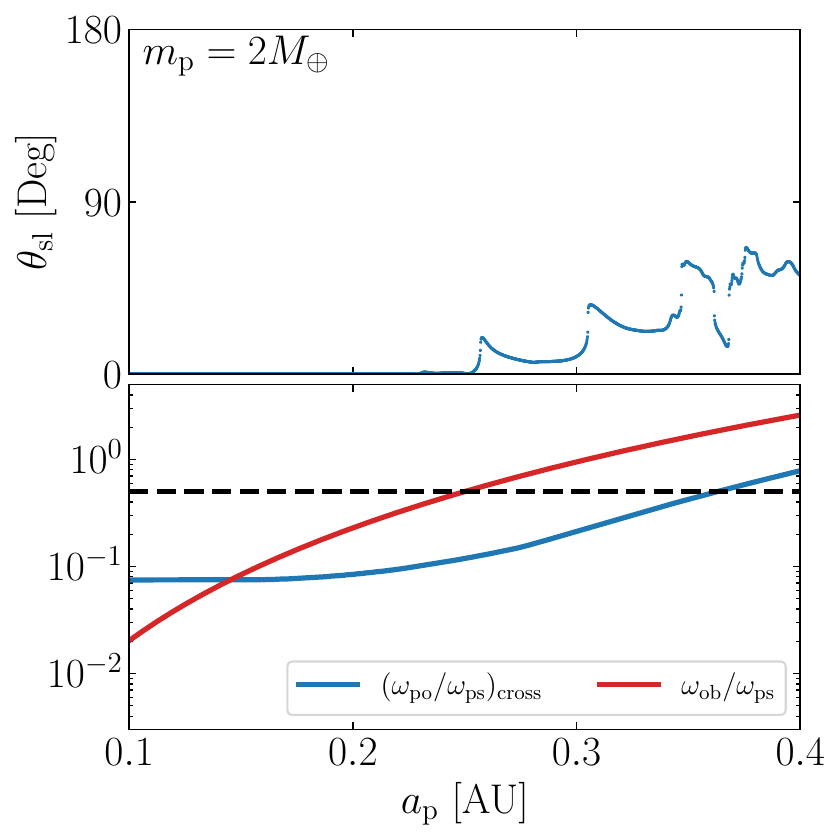}
    \includegraphics[width=0.76\columnwidth]{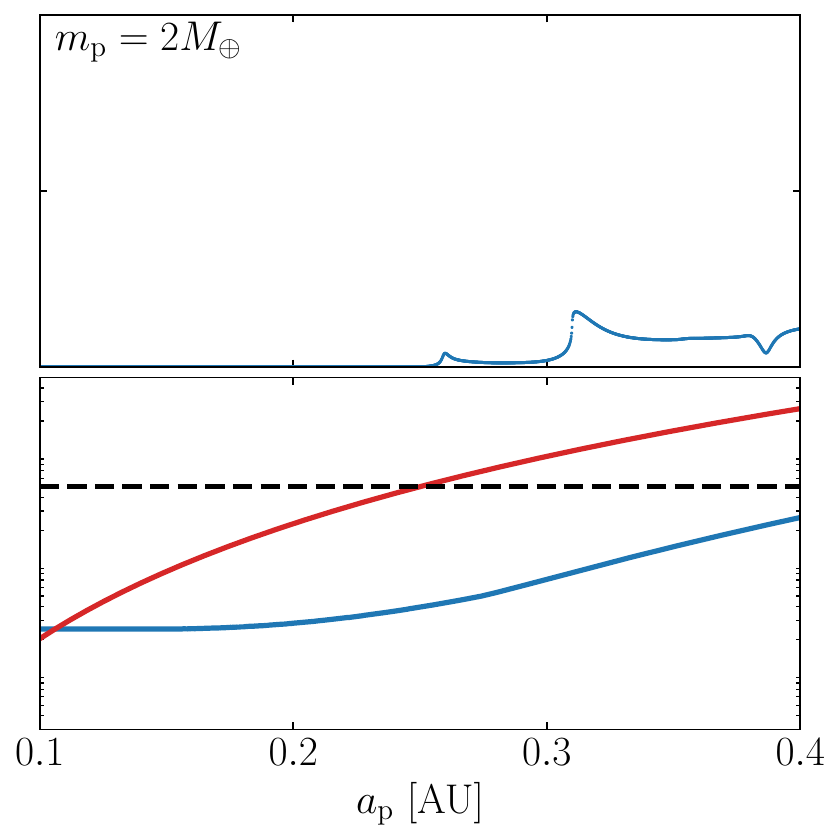}
    \caption{Same as the left panels of Fig.~\ref{fig:scanSEWJ} but for initial
    disk masses $0.3\times$ (left) and $0.1\times$ (right) as large. It can be
    seen that the transition to nonzero obliquities still occurs at the same
    semimajor axes due to the resonance crossing condition given by
    Eq.~\eqref{eq:2d_res_cross}, but the overall scale of the obliquities is
    smaller due to the weaker forcing strength given by
    Eq.~\eqref{eq:2d_res_jump}. }\label{fig:lighter}
\end{figure*}

On the other hand, if no precession resonance was crossed before the PPD has
opened a gap, then the analysis of the current section can be adapted, with
appropriately lower values for the inner and outer disk masses ($M_{\rm i, i}$
and $M_{\rm o, i}$).
We explore this effect by studying the evolution with disk masses $0.3\times$
and $0.1\times$ the values used in Section~\ref{ss:twop_analysis};
the results are shown in Fig.~\ref{fig:lighter}.
Smaller obliquities are observed because the resonant forcing strength $\propto
\omega_{\rm lo} / \omega_{\rm ls}$ is weaker (see the bottom panels).
Given the $\sim$ few Myr delay for photoevaporative gap opening, the
low obliquities obtained with these low disk masses are more likely to reflect
the misalignments in real systems.

\section{Summary and Discussion}\label{s:summary}

\subsection{Key Results}

In this paper, we have studied the disk dissipation-driven stellar obliquity
excitation mechanism in systems containing an oblate star, a protoplanetary disk
(PPD) with an embedded planet, and a moderately distant ($\sim
300\;\mathrm{au}$), inclined binary companion.
Compared to the related previous works (\citealp{lai2014star,
zanazzi2018Paper1}, see Section~\ref{s:intro}), we consider a range of planet
masses and semi-major axes, and we obtain analytical understanding and formulae
for the final stellar misalignments for single, idealized PPD models.
We also examine the effect of various non-ideal effects, such as a nonzero
star-to-disk angular momentum ratio and non-homologous disk dissipation.
Our main results are:
\begin{itemize}
    \item
        For the simplest disk evolution models, where the disk is treated as a
        rigid plate with a finite extent  and dissipates homologously,
        and assuming that the spin angular momentum of the star ($S$) is
        much smaller than that of the disk ($L$), we find that the resulting
        stellar obliquities can be computed analytically
        (Section~\ref{ss:th_sllL}) and are broadly distributed between $60^\circ$
        and $180^\circ$ (black dashed line in Fig.~\ref{fig:histall}).
        These obliquities result from the resonance crossing (and change to the
        phase-space topology) between the precession of the stellar spin driven
        by the ``disk + planet'' torque and the precession of the ``disk +
        planet'' system driven by the binary torque.

    \item
        At the next order of approximation, we show that the finite spin
        angular momentum of the star weakens the obliquity excitation when the
        disk is either well aligned ($\theta_{\rm lb, i} \approx 0^\circ$) or
        anti-aligned ($\theta_{\rm lb, i} \approx 180^\circ$) with the binary,
        or when they are nearly perpendicular ($\theta_{\rm lb, i} \approx
        90^\circ$).
        The size of the finite $S/L$ corrections depends on the angular momentum
        ratio at resonant excitation $(S / L)_{\rm c0}$, given by
        Eq.~\eqref{eq:def_slc}.
        Notably, the disk and binary become aligned for $\theta_{\rm lb, i}
        \lesssim \sqrt{(S / L)_{\rm c0}}$, and final obliquities with $\abs{\cos
        \theta_{\rm lb, f}} \lesssim (S / L)_{\rm c0}$ cannot be obtained
        (Fig.~\ref{fig:histall}).

        Furthermore, we find that while giant planets can be become misaligned
        for $a_{\rm p} \gtrsim 0.2\;\mathrm{au}$, small planets are readily
        misaligned as close in as $a_{\rm p} \gtrsim 0.03\;\mathrm{au}$
        (Section~\ref{ss:param_space}) for the fiducial disk parameters.

    \item
        When including a photoevaporatively opened gap in the PPD at
        $a \simeq 2\;\mathrm{au}$ in the disk evolution model, we find that
        planets embedded in the inner disk generally do not develop large
        misalignments with the stellar spin axis (Section~\ref{s:broken_disk}).

        In addition, we identify a new resonance (Eq.~\ref{eq:2d_res_cross}) by
        which planets with $a_{\rm p} \gtrsim 0.3\;\mathrm{au}$ may still become
        misaligned if the gap is opened sufficiently early
        (Section~\ref{ss:twop_analysis}).
        However, we also find that, for realistic gap opening times, this
        resonance is sufficiently weak, such that good alignment is expected
        (Section~\ref{ss:gap_timing}).
\end{itemize}
All the effects studied in this paper can shape the obliquity distributions of
stars with binary companions at early stages in their evolution.
We next discuss the limitations and applicability of our study.

\subsection{Discussion}\label{ss:discuss}

A general conclusion of our study is that, in the presence of binary companions,
planet-hosting stars can have significant ``primordial'' obliquities generated
in the PPD phase, especially for planets with semi-major axes $\gtrsim
1\;\mathrm{au}$.
Since the binary fraction of planet-hosting stars is currently estimated to be
$\sim 30\%$, and many of these binary companions orbit at separations of a few
hundreds of au \citep{horch2014_companions, matson2018_companions,
ziegler2020_companionsupp, colton2021_companions, offner2023review,
schlagenhauf2024_companions}, we expect the result of our study to have wide
applications.
% However, the semimajor axis distribution of field binaries peaks around $\sim
% 100\;\mathrm{au}$ binaries \citep{raghavan2010binary, offner2023review}, so the
% occurrence rate of the $\gtrsim 300\;\mathrm{au}$ binaries we study is lower
% than the field binary fraction.
% However, since these closer binaries truncate the protoplanetary disk
% (PPD) and suppress planet formation \citep{artymowicz1994disk, miranda2015disk},
% the wide binary fraction conditioned on planet hosts will somewhat compensate
% for this effect.

% Another complication of our mechanism is a lack of distant stellar-mass binary
% companions around hot Jupiters with large obliquities (\citealp{ngo2015}, but
% see also the other papers in the Friends of Hot Jupiters survey, e.g.\
% \citealp{knutson2014friends, piskorz2015friends}).
In fact, these statistics likely underestimate the applicability of our
mechanism.
The characteristic rate for strong encounters (those with minimum distance
$r_{\min} \lesssim 300\;\mathrm{au}$) in the stellar birth cluster is
\citep{rodet2021sulai}
\begin{align}
    \Gamma_{\rm close}
    \approx{}& \frac{1}{20\mathrm{Myr}}
        \p{\frac{n_\star}{10^3\;\mathrm{pc^{-3}}}}
        \p{\frac{M_{\rm tot}}{2M_\odot}}\nonumber\\
        &\times \p{\frac{r_{\min}}{300\;\mathrm{au}}}
        \p{\frac{\sigma_\star}{1\;\mathrm{km/s}}},
\end{align}
where we have taken fiducial parameters for $n_\star$, the stellar density, and
$\sigma_\star$, the velocity dispersion, from the birth cluster of the Solar
System \citep[e.g.][]{pfalzner2013birthcluster} and the Scorpius-Centaurus OB
association \citep{wright2018}.
Since characteristic birth cluster lifetimes are also $\sim 20\;\mathrm{Myr}$
\citep{pfalzner2013birthcluster}, it is quite possible for the binary companions
in our study to be unbound after the stellar obliquity is excited in the first
few Myr.
This is in agreement with the high observed binary fraction of young stars
\citep{goodwin2005_binary, offner2023review} and with the predicted evolution of
the binary fraction from N-body simulations of star-forming regions
\citep[e.g.][]{kroupa1995binary, thies2015_binfrac}.
Our analysis does not account for the stochastic wandering of the binary plane
due to weak encounters.
It's conceivable that the effect of such weak encounters is a broadening of the
obliquity distributions compared to our fiducial results, and further study is
required to quantify this effect.

Some of the systems studied in this paper contain disks that are substantially
misaligned with their distant binary companions.
Such disks may undergo von Zeipel-Lidov-Kozai (ZLK) oscillations
\citep{martin2014ZKLdisk, fu2015diska}, which can have critical minimum
inclinations smaller than the standard $39^\circ$ for select regions
of parameter space \citep{zanazzi2017diskLK, lubow2017diskLK}.
For our fiducial parameters, ZLK disk oscillations are likely to be suppressed by
both the disk's self-gravity \citep{fu2015diskb, zanazzi2018Paper2} when
\begin{equation}
    M_{\rm d} \gtrsim 0.005 M_{\rm b}\p{\frac{6r_{\rm out}}{a_{\rm b}}}^3,
\end{equation}
and by the disk's pressure gradients when \citep{zanazzi2018Paper2}
\begin{equation}
    a_{\rm b} \gtrsim 4.2r_{\rm out} \p{\frac{M_{\rm b}}{M_\star}}^{1/3}
        \p{\frac{h_{\rm out}}{0.1}}^{-2/3},
\end{equation}
where $h_{\rm out} = H(r_{\rm out}) / r_{\rm out}$ is the disk aspect ratio at
the outer edge.
This suggests that ZLK oscillations will only set in after the obliquity
excitation, which typically occurs at a disk mass of $\sim 0.01M_\star$
(Eq.~\ref{eq:eta_i}).
Even if the disks somehow manage to undergo ZLK oscillations, they would
experience shocks during high-eccentricity phases and quickly settle to the edge
of the ZLK window, effectively described as a change to the distribution of the
initial $\theta_{\rm lb, i}$ values.

As a last dynamical point, we have considered only a single embedded planet in
this work.
Since many planetary systems have multiple planets, we briefly comment on recent
work investigating the evolution of multiplanetary systems in the presence of a
photoevaporating disk (without a stellar binary companion).
Previously, \citet{spalding2020} considered the effect of a giant planet and an
inner planetary system embedded in a dissipating disk surrounding a misaligned
star.
They found that, as long as the disk dissipates sufficiently slowly, the inner
planets become aligned with the stellar equator due to a Laplace plane
transition; their system architecture resembles that considered in our
Section~\ref{ss:double}.
However, we found that planetary systems can readily attain misalignment:
the difference is that our planetary system is not aligned with the perturber
(binary companion) while they considered an aligned gas giant perturber (recall
that the final stellar obliquity is equal to the initial disk-perturber
misalignment, Section~\ref{ss:double}).
This scenario was further extended by \citet{fu2024_diskbifurc}, who found that
a dramatic saddle node bifurcation (similar to the one studied in this work) at
sufficiently large stellar obliquities can misalign inner planets from both
exterior planets and the stellar equator.
Taken together, it is clear that multiplanetary systems can introduce additional
secular resonances that misalign the planets' orbits.

In comparison to observations, recent works suggest that the distribution of
misalignment angles between the orbits of circumprimary planets and those of
not-so-distant binary companions ($\lesssim 400\;\mathrm{au}$) modestly favors
alignment compared to an isotropic distribution \citep{dupuy2022orbital,
christian2022align, lester2023align}, though this trend may be predominantly
driven by small planets \citep{behmard2022align, christian2024small}. The
stellar spins in such systems may also be preferentially aligned with both
orbital planes \citep{rice2023qatar6, rice2024spinorbitorbit}. We have found
that disk-driven obliquity excitation also drives alignment of the disk and
binary planes due to angular momentum constraints (Figs.~\ref{fig:tau3plots}
and~\ref{fig:histall}), an effect that was first identified by
\citet{anderson2018teeter}, though the binary companions we consider are
generally more compact than those in the \citet{rice2024spinorbitorbit} sample.
Our mechanism is nevertheless incompatible with the spin-orbit-orbit aligned
sample, which may be the result of additional dissipative processes
\citep{gerbig2024} or simply good alignment at formation.

We have assumed that transition disks form via a photoevaporated gap at $\sim
2\;\mathrm{au}$, hydrodynamically separating the inner and outer disks.
Such systems may be observed as ``dippers'', which consist of double-disked
systems that have a broad mutual inclination distribution
\citep{ansdell2020dipper}.
This picture may change for A dwarfs and above ($\gtrsim 1.5M_{\odot}$), where
studies suggest that photoevaporation is significantly less efficient
\citep{nakatani2023adwarf}, and the traditional picture of inside-out clearing
may be replaced by similar dissipation rates in the inner and outer regions of
the disk \citep{ronco2024IMstar}. Other studies also find a mass dependence of
the disk evolution \citep{komaki2023ppd}. This is suggestive of a temperature
dependence of the mechanism studied in this paper, which may result in
larger obliquities for hot stars.
Separately, MHD winds \citep{suzuki2016mhdwind, kunitomo2020mhd} may suppress
the formation of such a gap, and gap-opening planets
\citep{duffell2015transition, dong2016transition, picogna2023transition} may
introduce additional gravitational coupling in the two-disk system if they are
responsible for the opened gap.
The stellar obliquity evolution when considering all these confounding factors
is hard to estimate concisely, but a rich variety of outcomes is clearly
possible.

There is emerging evidence that warm giant planets exhibit low ($\lesssim
30^\circ$) obliquities via the Stellar Obliquities in Long-period Exoplanet
Systems (SOLES) survey \citep[e.g.][]{rice2021soles1, wang2022soles2}
tentatively attributed to coplanar high-eccentricity migration
\citep{petrovich2015, radzom2024soles}.
This scenario requires a small primordial obliquity which is retained during the
dynamical migration process.
However, when the homologous disk models are used, cold giants in systems with
binary companions are expected to be born with large obliquities
(Fig.~\ref{fig:param_space}).
Our broken disk model yields well-aligned giant planets up to the phoevaporative
gap radius of $\sim 1\;\mathrm{au}$ (Section~\ref{s:broken_disk}).
This may be required to achieve agreement with observations if the primordial
stellar binary fraction is indeed large at early times.

Finally, in the obliquity excitation mechanism considered in this paper, the
angular momentum evolution of the central star driven by the disk must be
dominated by gravitational, rather than hydrodynamical (accretion), coupling.
Such a protostellar system is typically denoted as a Class II young stellar
object, for which typical disk mass estimates (e.g.\ $M_{\rm d} \simeq
0.01M_\star$, \citealp{williams2011_diskreview}) may be lower than the fiducial
value taken in our paper.
However, as long as the spin-orbit resonance is crossed, which requires that
Eq.~\ref{eq:eta_i} (which also depends on the binary companion's properties) be
less than unity initially, the specific value of the initial disk mass does not
significantly affect our results.
Moreover, the total masses of disks are difficult to constrain precisely, since
they are dominated by difficult-to-observe components such as cold H$_2$, and
more easily observable tracers such as disk dust mass sometimes underestimate
the disk gas mass \citep[e.g.][]{facchini2019_gasdustdisk,
manara2023_ppv7_disk}.

\section{Acknowledgements}

We thank Justin Tan for assistance with the initial formulation of this work
and the anonymous referee whose careful review significantly
improved the clarity of this work.
YS thanks Simon Albrecht, Ilse Cleeves, Mark Dodici, Konstantin Gerbig, Kaitlin
Kratter, Steven Lubow, Masahiro Ogihara, Eve Ostriker, Jared Siegel, Yanqin Wu,
Joshua Winn, and J.\ J.\ Zanazzi for useful comments and fruitful discussions.
This work has been supported in part by NSF grant AST-2107796 and by NASA
FINESST grant 19-ASTRO19-0041.
YS is supported by a Lyman Spitzer, Jr. Postdoctoral Fellowship at Princeton
University.

\section{Data Availability}

The data referenced in this article will be shared upon reasonable request to
the corresponding author.

\bibliography{Su_diskstarobl}
\bibliographystyle{aasjournal}

\appendix

\section{Disk Precession Frequencies}\label{app:wio}

In this paper, three nontrivial precessional effects involve rigid disks: when
the disk drives precession of an inner body (e.g.\ $\omega_{\rm sd}$,
Eq.~\ref{eq:def_wsd}), when an outer body drives precession of an inner disk
(e.g.\ $\omega_{\rm db}$, Eq.~\ref{eq:def_wdb}), or when an outer disk drives
precession of an inner disk (e.g.\ $\omega_{\rm io}$, Eq.~\ref{eq:def_wio}). The
precession rate in the first case can be calculated by simply integrating over
the disk as a series of rigid, thin rings. The precession rate in the second
case is obtained by evaluating the total torque on the disk as shown in
\citet{lai2014star}. The final case is obtained by a combination of these two
procedures and is shown below:
\begin{align}
    \bm{T}_{\rm io} ={}& \omega_{\rm io}
        \p{\uv{l}_{\rm o} \cdot \uv{l}_{\rm i}}
        \p{\uv{l}_{\rm o} \times \bm{l}_{\rm i}}\nonumber\\
        ={}& \frac{3GM_{\rm i}M_{\rm o}}{4}
                \p{\uv{l}_{\rm o} \cdot \uv{l}_{\rm i}}
            \int\limits_{0}^{r_{\rm i, out}}
                \frac{r_{\rm in}^2}{r_{\rm i, out}}\;\mathrm{d}r_{\rm in}
            \int\limits_{r_{\rm o, in}}^\infty
                \frac{1}{r_{\rm out}^3 r_{\rm o, out}}\;\mathrm{d}r_{\rm out}
                \nonumber\\
        ={}& \frac{GM_{\rm i}M_{\rm o}}{4}
                \p{\uv{l}_{\rm o} \cdot \uv{l}_{\rm i}}
                \p{\uv{l}_{\rm o} \times \uv{l}_{\rm i}}
            \frac{r_{\rm i, out}^2}{2r_{\rm o, out}r_{\rm o, in}^2},\\
    \omega_{\rm io} ={}& T_{\rm io} \s{\frac{2}{3}M_{\rm i}\sqrt{GM_\star r_{\rm
            i, out}}}^{-1}\nonumber\\
        ={}& \frac{3}{16}\frac{M_{\rm o}}{M_\star}
            \sqrt{\frac{GM_\star}{r_{\rm i, out}^3}}
            \frac{r_{\rm i, out}^3}{r_{\rm o, out} r_{\rm o, in}^2}.
\end{align}

\end{document}